\documentclass[a4paper,11pt]{article}
\usepackage{a4wide}
\usepackage{graphicx}
\usepackage{subcaption}
\usepackage{xcolor}
\usepackage{amsmath,amsfonts,amssymb,amstext,graphicx}
\usepackage{placeins}
\usepackage[colorlinks=true,  citecolor=blue, linkcolor=blue, urlcolor=black]{hyperref}
\usepackage[numbers,sort&compress]{natbib}
\usepackage{breakurl}

\usepackage{cancel}
\usepackage{empheq}
\usepackage{float} 
\usepackage{subcaption}
\usepackage{authblk}
\usepackage{comment}
\usepackage{blindtext}
\usepackage{cases}
\usepackage{ulem}

\usepackage[bottom]{footmisc}
\usepackage{lineno}

\usepackage{tikz}
\usetikzlibrary{arrows, positioning}

\newcommand{\dv}[2]{\frac{d#1}{d#2}}

\newcommand{\com}[2]{\big{[}#1,#2\big{]} }

\newcommand{\trd}[2]{{{\rm Tr}_{#1} {#2}}}

\newcommand{\av}[1]{\langle #1 \rangle}

\begin{document}  
\title{\centerline \textbf {Cross-Spectral Reservoir Correlations as a Resource for Finite-Time Quantum Otto Engines}}
\author[1]{Siddhartha Dutta
\thanks{siddhartha.dutta.phy23@gm.rkmvu.ac.in}}

\author[1]{Sujay Mondal 
\thanks{sujay.mondal.phy23@gm.rkmvu.ac.in}} 

\author[1]{Ankush Das
\thanks{ankushdasself@gmail.com}} 

\author[2,3]{Anumita Mukhopadhyay 
\thanks{anumitamukherjee455@gmail.com}} 

\author[1]{Abhijit Bandyopadhyay%
\thanks{abhijit.phy@gm.rkmvu.ac.in}}

\affil[1]{Department of Physics, Ramakrishna Mission Vivekananda
Educational and Research Institute,
Belur Math, Howrah 711202, West Bengal, India} 

\affil[2]{Center for Quantum Engineering, Research and Education (CQuERE),
TCG CREST, Salt Lake, Sector 5, Kolkata 700091, India.}

\affil[3]{Academy of Scientific and Innovative Research (AcSIR), Ghaziabad- 201002, India.}
\date{\today}

\maketitle

 \begin{abstract} 
We investigate the thermodynamic consequences of longitudinal--transverse cross-spectral reservoir correlations in a finite-time quantum Otto engine with a two-level working medium. Each reservoir couples through excitation--relaxation and dephasing channels whose fluctuations are characterized by a Hermitian positive-semidefinite spectral-density matrix, with the off-diagonal elements encoding their cross correlations. The finite-time isochoric dynamics is derived within the second-order time-convolutionless framework, without imposing the Markov limit at the outset, so that finite reservoir-memory effects can enter through time-dependent dissipative and reservoir-induced coherent contributions. The resulting dynamics is then recast in Bloch-vector form to construct the stroke-resolved cycle dynamics. At fixed auto-spectral densities, cross-spectral correlations modify the populations and coherences of the working medium and thereby its thermodynamic performance. Increasing the correlation strength can enhance the output power, with the enhancement controlled by the cross-spectral phase and characteristic frequency scale. The correlations also reshape the transient cycle-to-cycle evolution and the approach to periodic operation, while the limit-cycle efficiency remains fixed at the Otto value for the population-preserving unitary strokes considered here. These results establish off-diagonal reservoir spectra as an additional resource for controlling finite-time quantum thermal machines. 
\end{abstract}

\section{Introduction}
\label{sec:introduction} 
The extension of thermodynamics to microscopic quantum systems has established
quantum thermal machines as a natural platform for investigating energy
conversion in regimes where coherence, correlations, discrete spectra, and
environmental backaction can influence thermodynamic performance
\cite{Alicki1979,Quan2007,Kosloff2013}. Among cyclic protocols, the quantum
Otto cycle, comprising two unitary work strokes interspersed with two isochoric
heat-exchange strokes, provides a particularly simple and convenient framework for such
studies \cite{Kieu2004,Quan2007}. Its minimal realization with a two-level
working medium has also been demonstrated experimentally using a spin-$\frac12$
system \cite{Peterson2019}. At finite cycle times, incomplete thermalization
during the isochores and coherence generated by nonadiabatic driving can
substantially affect work extraction, power, and irreversibility
\cite{Brandner2017,Camati2019}. The performance of a finite-time quantum Otto
engine is therefore governed not only by the working-medium spectrum and
reservoir temperatures, but also by the open-system dynamics during the
heat-exchange strokes.\\

This observation has stimulated considerable interest in using the environment
itself as a thermodynamic resource. 
Squeezed thermal reservoirs, whose fluctuations are redistributed asymmetrically between
conjugate quadratures, can modify the power and efficiency of quantum Otto
engines \cite{Rossnagel2014}. Effective negative-temperature
reservoirs characterized by population inversion, can enable operating
regimes unavailable to conventional positive-temperature baths
\cite{deAssis2019,Mendonca2020}. 
Strong system--reservoir coupling effects become thermodynamically important
beyond the conventional weak-coupling regime, where interaction and decoupling
energies may contribute appreciably to the thermodynamic bookkeeping \cite{Newman2017}. 
Reservoir memory provides a complementary route beyond the Markovian limit:
finite-time energy backflow can enhance work extraction in a quantum Otto
engine \cite{Shirai2021}, while non-Markovian thermal operations can increase
the work output at fixed efficiency and suppress work fluctuations relative
to their Markovian counterparts \cite{Ptaszynski2022}. 
These results show that reservoir temperature alone does not fully determine
the performance of a quantum thermal machine; the environmental spectral
structure, memory, and correlations can provide additional thermodynamic
control parameters.\\

An additional environmental effect arises when a two-level working medium
couples to a common reservoir through multiple system operators. The resulting
transverse and longitudinal fluctuations, associated with
excitation--relaxation and dephasing, are often treated as independent, although
common environmental degrees of freedom can generate finite cross correlations.
A frequency-resolved description then requires a matrix-valued spectral
density, whose diagonal elements characterize the individual noise spectra and
whose off-diagonal elements encode the cross-spectral response.   
Such correlated multichannel noise is relevant across superconducting,
semiconductor, and defect-based qubits, as well as in systems coupled to
phononic and electromagnetic environments
\cite{Makhlin2001,Ithier2005,Paladino2014,Witzel2006,Krummheuer2002,
McCutcheonNazir2010,NazirSchaller2018}. Such correlations generated by common
environmental degrees of freedom can modify decoherence and dissipative
dynamics beyond predictions based on independent noise channels
\cite{Palma1996,FicekTanas2002,Jeske2013}.
 Their experimental relevance is now
well established:
these correlations are, in principle, both measurable and can be engineered
with present open-system control techniques. Quantum-noise spectroscopy uses
controlled qubit evolution as a frequency-selective filter and can separate
auto- and cross-spectral contributions associated with different noise
channels \cite{PazSilva2019,Khan2024}; related protocols have already extracted
the magnitude and sign of correlated fluctuations experimentally
\cite{Gustavsson2011,vonLupke2020,RojasArias2026}. From the reservoir-engineering
side, programmable bosonic environments with controllable temperatures and
spectral densities have recently been realized in trapped-ion simulations of
spin--boson models \cite{Sun2025}, while artificial reservoirs with tunable
thermodynamic properties have been implemented in nuclear-spin platforms
\cite{Mendonca2020}. Microscopically, if a common bosonic mode $k$ couples to
the transverse $(x)$ and longitudinal $(z)$  operators with amplitudes $g_{xk}$ and
$g_{zk}$, respectively, the corresponding cross spectrum contains products of
the form $g_{xk}g_{zk}^{*}$. Independent control of the coupling amplitudes and
their relative phases therefore provides a direct route to tailoring the
magnitude and phase of the cross-spectral response, subject to the positivity
constraints of the spectral-density matrix. More generally, correlated
stochastic drives, common-mode fluctuations, and programmable structured
bosonic modes provide experimentally accessible routes for synthesizing such
effective environments. 
These results establish off-diagonal environmental 
spectra as physically
meaningful and experimentally accessible quantities, and suggest that
cross-spectral correlations may serve as a distinct reservoir-engineering
resource alongside temperature, squeezing, and the diagonal spectral response.\\

A recent analysis within the second-order time-convolutionless framework has
shown that cross-spectral correlations can produce nonmonotonic relaxation,
transient coherence revival, and intervals of suppressed relaxation, with the
cross-spectral amplitude, phase, bandwidth, and delay providing distinct
dynamical control parameters \cite{Dutta2026}. These results motivate a natural
thermodynamic question: how are such correlation-induced modifications of the
dissipative dynamics transferred to the heat, work, power, and efficiency of a
cyclic quantum engine? 
Despite extensive studies of engineered, structured, and non-Markovian
reservoirs, the thermodynamic implications of frequency-resolved correlations
between distinct coupling channels remain comparatively less explored.\\

Motivated by these considerations, we investigate a finite-time quantum Otto
engine with a two-level working medium, undergoing two unitary work strokes
interspersed with hot and cold isochoric strokes during which it is alternately
coupled to independent bosonic reservoirs. While the two reservoirs remain
mutually uncorrelated, each couples to the working medium through transverse and
longitudinal channels associated with excitation--relaxation and dephasing,
respectively. The corresponding fluctuations are described by a Hermitian
positive-semidefinite spectral-density matrix, with real diagonal elements defining
the individual auto-spectra and complex off-diagonal   elements encoding the
longitudinal--transverse cross correlations. We parametrize the latter through
a complex correlation coefficient, whose amplitude and phase control the
strength and relative phase of the cross-spectral response while preserving the
positivity of the spectral matrix.  
We derive the finite-time reduced dynamics 
of the working medium 
during each isochoric stroke to second order
in the working-medium--reservoir coupling, retaining both dissipative and
Lamb-shift contributions. Since the Markov limit is not taken in deriving the isochoric propagators, the TCL2 coefficients retain the finite-time reservoir correlation functions. The resulting reduced dynamics is therefore generally non-Markovian at finite
isochore times, with the Markovian dynamics recovered only when the reservoir correlation time is much shorter than the relevant system and stroke timescales. The cross-spectral terms thereby  modify the evolution of populations and coherences during the isochores and, consequently,
the state entering the subsequent work stroke.  
To isolate this effect, we
compare correlated and uncorrelated reservoirs at fixed auto-spectral densities
while varying the cross-spectral amplitude and phase, and examine the resulting
changes in heat exchange, cycle work, power, and efficiency. Since correlations
in the hot and cold reservoirs prepare the states entering the expansion and
compression strokes, respectively, they provide distinct routes through which
the off-diagonal reservoir spectrum controls the finite-time thermodynamic
performance of the engine.\\

In this work, we perform a quantitative analysis of how
longitudinal--transverse cross-spectral correlations influence the finite-time
thermodynamic performance of a quantum Otto engine beyond conventional control
through the reservoir temperatures and the individual auto-spectral densities.
We identify parameter regimes in which suitably engineered cross-spectral
correlations improve the engine performance relative to the corresponding case
of statistically independent longitudinal and transverse coupling channels.
We find that,
at fixed auto-spectral densities,   increasing the cross-spectral correlation
strength enhances the generated power, with the extent of the enhancement
controlled by both the cross-spectral phase and its characteristic frequency
scale. The cross-correlations also modify the cycle-to-cycle evolution of the
efficiency by changing the rate at which the engine approaches its asymptotic
Otto value. These results identify the off-diagonal components of the reservoir  
spectral-density matrix as an independent resource for controlling finite-time
quantum-engine performance.\\

The remainder of this article is organized as follows. In Sec.\ \ref{sec:model}, we
introduce the model, reduced dynamics, and thermodynamic framework of the
finite-time quantum Otto engine. Sec.\ \ref{sec:micmodel}  specifies the microscopic
model of the working medium and reservoirs, Sec.\ \ref{sec:tcl2}  develops the TCL2
reduced dynamics during the isochoric strokes, including the contributions
of the reservoir auto- and cross-spectral densities, and  Sec.\ \ref{sec:bkp}
establishes the weak-coupling thermodynamic bookkeeping for heat, work,
efficiency, and power. In Sec.\ \ref{sec:bloch1}, we recast the isochoric TCL2 dynamics in
Bloch-vector form, providing a compact description of the population and
coherence dynamics. Sec.\ \ref{sec:bloch2} develops the Bloch-vector framework for the
finite-time thermodynamic analysis, including the stroke-resolved propagation,
cycle-to-cycle evolution, and transient and limit-cycle thermodynamics.
In Sec.\ \ref{sec:results}, we present and discuss the numerical results, with emphasis on the
dependence of the engine performance on the strength, phase, and characteristic
frequency scale of the reservoir cross-spectral correlations. Finally,
Sec.\ \ref{sec:conclusion} summarizes our main conclusions.

\section{Model, Dynamics, and Thermodynamics of a Quantum Otto Engine} 
\label{sec:model}
 \begin{figure}[t]
\centering
\includegraphics[width=0.8\linewidth]{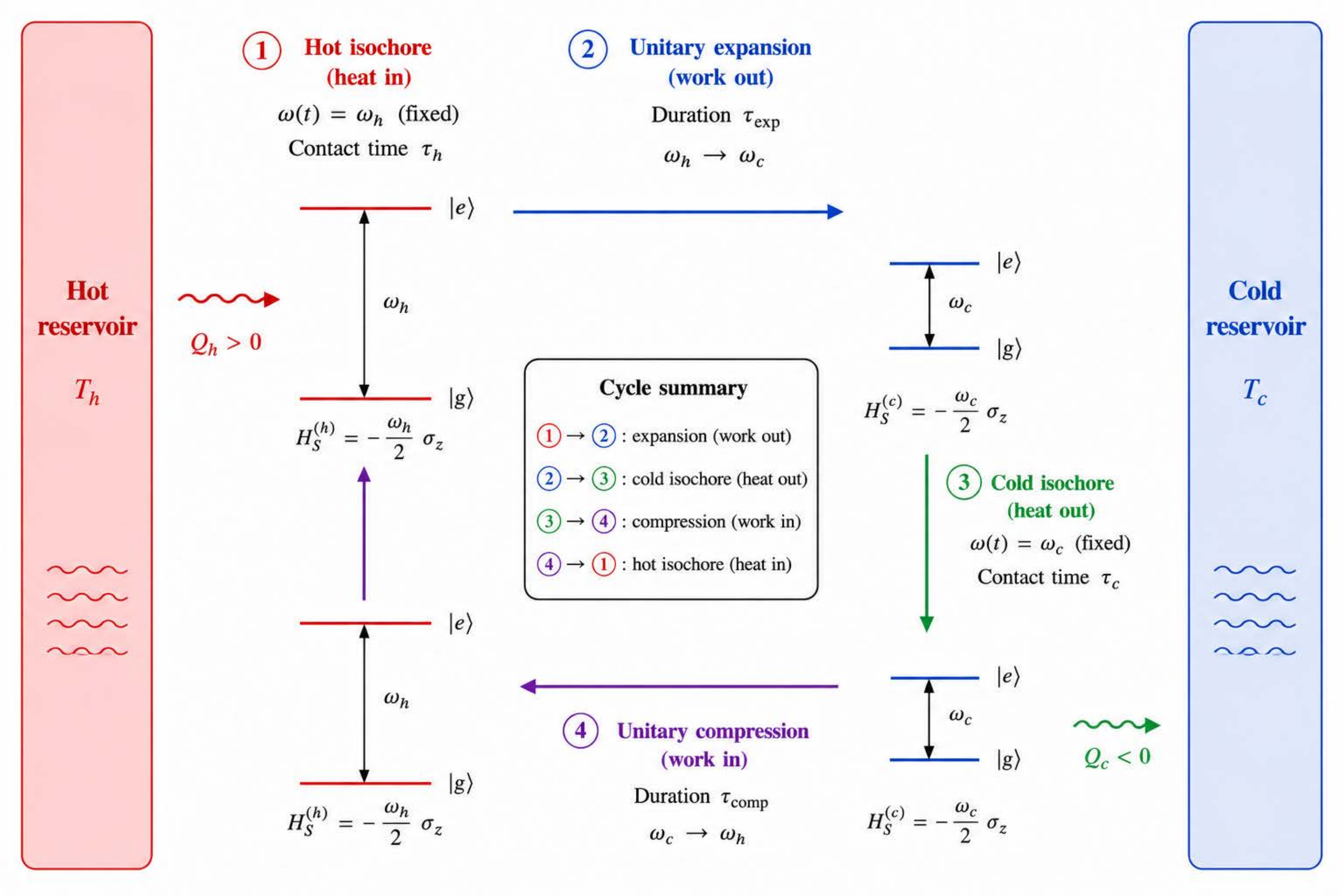}
\caption{Illustration of a finite-time quantum Otto cycle:The two-level working medium $S$ exchanges heat with the hot and cold reservoirs during the isochoric strokes, while the energy-level spacing of its bare Hamiltonian $H_S$ is held fixed at $\omega_h$ and $\omega_c$, respectively, with $\omega_h>\omega_c$. Here, $Q_h$ denotes the heat absorbed from the hot reservoir, whereas $Q_c$ denotes the heat released to the cold reservoir. During each isochore, the system--reservoir coupling also induces a Lamb-shift correction $H_{\rm LS}^{(\alpha)}(t)$ to the bare working-medium Hamiltonian. The isolated expansion and compression strokes vary the bare level spacing as $\omega_h\rightarrow\omega_c$ and $\omega_c\rightarrow\omega_h$, respectively.}
\label{fig:1}
\end{figure}
We consider a two-level system $S$ as the working medium of a finite-time
quantum Otto engine operating between independent hot ($h$) and cold ($c$)
bosonic reservoirs \cite{Kieu2004,Quan2007,Kosloff2013,Peterson2019}
through two unitary driving strokes interspersed with two isochoric
heat-exchange strokes, as illustrated in Fig.~\ref{fig:1}.
The cycle follows the sequence:
$\text{hot isochore} 
\to  \text{unitary expansion}
\to  \text{cold isochore}  
\to  \text{unitary compression}$,
after which the working medium returns to its initial Hamiltonian configuration and the same sequence is repeated in subsequent cycles.

\subsection{Microscopic Model of the Working Medium and Reservoirs}
\label{sec:micmodel}
The working medium is described by the bare Hamiltonian
\begin{eqnarray}
H_S(t)
&=&
-\frac{\omega(t)}{2}\sigma_z .
\label{eq:HSbare}
\end{eqnarray}
During the unitary work strokes, the system is decoupled from both reservoirs
and the level spacing is externally modulated from $\omega_h$ to $\omega_c$
($\omega_h>\omega_c$) during expansion and from $\omega_c$ back to $\omega_h$
during compression. 
 During isochore $\alpha$ ($\alpha=h,c$), the working medium is coupled
to the corresponding reservoir at temperature $T_\alpha$, while the
external control is held fixed so that the level spacing remains
constant at $\omega_\alpha$ yielding the bare system Hamiltonian of
the form
\begin{eqnarray}
H_S^{(\alpha)}
&=&
-\frac{\omega_\alpha}{2}\sigma_z,
\qquad
\alpha=h,c\,,
\label{eq:HSalpha}
\end{eqnarray}

Each reservoir $\alpha=h,c$ is modeled as an independent bosonic bath with free Hamiltonian
\begin{eqnarray}
H_B^{(\alpha)}
&=&
\sum_k \omega_k^{(\alpha)}
b_{k,\alpha}^{\dagger} b_{k,\alpha},
\label{eq:HBalpha}
\end{eqnarray}
where $b_{k,\alpha}^{\dagger}$ ($b_{k,\alpha}$) creates (annihilates) a boson of frequency $\omega_k^{(\alpha)}$ in mode $k$ of reservoir $\alpha$
satisfying the canonical
bosonic commutation relations
\begin{eqnarray}
\left[
b_{k,\alpha},
b_{k',\alpha'}^{\dagger}
\right]
&=&
\delta_{kk'}\delta_{\alpha\alpha'},
\quad 
\left[
b_{k,\alpha},
b_{k',\alpha'}
\right]
=
\left[
b_{k,\alpha}^{\dagger},
b_{k',\alpha'}^{\dagger}
\right]
=
0 .
\label{eq:bosonic_comm}
\end{eqnarray}
The reservoir $\alpha$  
prepared in thermal equilibrium at temperature $T_\alpha$ are described by
the Gibbs state $\rho_B^{(\alpha)}
= e^{-\beta_\alpha H_B^{(\alpha)}}/Z_\alpha $,
with $
Z_\alpha= {\rm Tr}_B [ e^{-\beta_\alpha H_B^{(\alpha)}}]$ and $\beta_\alpha=1/T_\alpha$.
Since
$[\rho_B^{(\alpha)},H_B^{(\alpha)}]=0$, the reservoir state is stationary
under the free evolution generated by $H_B^{(\alpha)}$. 
During isochore $\alpha$, the working medium couples to the reservoir
through transverse and longitudinal channels described by the interaction
Hamiltonian
\begin{eqnarray}
H_{SB}^{(\alpha)}
&=&
\sigma_x\otimes B_x^{(\alpha)}
+
\sigma_z\otimes B_z^{(\alpha)}\,,
\label{eq:HSBalpha}
\end{eqnarray}
where, $B_x^{(\alpha)}$ and $B_z^{(\alpha)}$ are reservoir operators associated
with the transverse ($\sigma_x$) and longitudinal ($\sigma_z$) coupling
channels, respectively. For the bare working-medium Hamiltonian proportional
to $\sigma_z$, the transverse channel induces excitation--relaxation
processes, whereas the longitudinal channel gives rise to dephasing.
Hermiticity of $H_{SB}^{(\alpha)}$ requires the corresponding reservoir
operators to be Hermitian: 
$\big[B_\mu^{(\alpha)}\big]^\dagger=
B_\mu^{(\alpha)}$, $\mu=x,z$. 
We consider reservoir operators that depend linearly on the bosonic creation
and annihilation operators,
\begin{eqnarray}
B_{\mu}^{(\alpha)}
&=&
\sum_k
\left[
g_{k\mu}^{(\alpha)} b_{k,\alpha}
+
\big(g_{k\mu}^{(\alpha)}\big)^*
b_{k,\alpha}^{\dagger}
\right],
\qquad
\mu=x,z ,
\label{eq:BmuAlpha}
\end{eqnarray}
where $g_{k\mu}^{(\alpha)}$ denotes the coupling amplitude between the
$\mu$-th system channel and the $k$-th mode of reservoir $\alpha$. Such linear
couplings arise naturally when a quantum system interacts with weak
fluctuations of phononic, electromagnetic, or other approximately harmonic
environmental fields~\cite{FeynmanVernon1963,CaldeiraLeggett1983,Breuer2002}.
Since $B_x^{(\alpha)}$ and $B_z^{(\alpha)}$ belong to the same reservoir,
the two coupling channels may interact with a common set of reservoir modes,
and the corresponding coupling amplitudes
$\{g_{kx}^{(\alpha)}\}$ and $\{g_{kz}^{(\alpha)}\}$ need not be orthogonal.
A nonvanishing overlap between these amplitudes then gives rise to cross
correlations between the longitudinal and transverse fluctuations of
reservoir $\alpha$~\cite{PazSilva2019,Dutta2026}.\\

\subsection{Isochoric TCL2 Dynamics of the Working Medium Induced by Reservoir Spectral Correlations}
\label{sec:tcl2}
The isochoric dynamics is described using a weak-coupling TCL2 master-equation approach \cite{BreuerPetruccione2002,RivasHuelgaPlenio2014,deVegaAlonso2017}, in which the Born expansion is truncated at second order while the finite-time
reservoir-memory integrals are retained. The generator is therefore time
local, but its coefficients remain time dependent and are determined by bath
correlation functions. The resulting finite-time dynamics is generally non-Markovian, with the Markovian master equation recovered only after the
additional long-time approximation.\\

During the isochores, the system-reservoir composite state
$\rho_{SB}^{(\alpha)}$ undergoes unitary evolution
generated by the Hamiltonian $(H_S^{(\alpha)}+H_B^{(\alpha)}+H_{SB}^{(\alpha)})$.
We formulate the system--reservoir dynamics in the interaction picture
with respect to the free Hamiltonian
$H_0^{(\alpha)}=H_S^{(\alpha)}+H_B^{(\alpha)}$,
where  the   unitary evolution
of the system-reservoir joint 
state $\widetilde\rho_{SB}(t) = e^{iH_0^{(\alpha)}t} \rho_{SB}(t)e^{-iH_0^{(\alpha)}t}$ is 
given by  
\begin{eqnarray}
\dv{}{t} \rho_{SB}^{(\alpha)}(t)
= -i \com{\widetilde H_{SB}^{(\alpha)}(t)}{\rho_{SB}^{(\alpha)}(t)}\,,
\label{eq:intpicevol1}
\end{eqnarray}
where  $\widetilde H_{SB}(t)=e^{iH_0t}H_{SB}e^{-iH_0t}$ represents the interaction Hamiltonian
in interaction picure. Using the form   \eqref{eq:HSBalpha},
 $\widetilde H_{SB}(t)$ can be expressed as
\begin{eqnarray}
\widetilde H_{SB}(t)
&=&
\sum_{\mu=x,z}
\widetilde{\sigma}_{\mu}^{(\alpha)}(t)
\otimes
\widetilde B_{\mu}^{(\alpha)}(t),
\label{eq:Hsbintpic}
\end{eqnarray}
where
$\widetilde{\sigma}_{\mu}^{(\alpha)}(t)
=e^{iH_S^{(\alpha)}t}\sigma_\mu e^{-iH_S^{(\alpha)}t}$
and
$\widetilde B_{\mu}^{(\alpha)}(t)
=e^{iH_B^{(\alpha)}t}B_\mu e^{-iH_B^{(\alpha)}t}$
are the corresponding interaction-picture system and reservoir
operators.
At the beginning of each isochoric stroke, we assume the working medium
and reservoir $\alpha$ to be uncorrelated,
$\rho_{SB}^{(\alpha)}(0)=\rho_S(0)\otimes\rho_B^{(\alpha)}$. The reservoir response
entering the reduced isochoric dynamics is encoded in the two-time
correlation functions
\begin{eqnarray}
C_{\mu\nu}^{(\alpha)}(t)
&=&
{\rm Tr}_{B}
\left[
\widetilde B_{\mu}^{(\alpha)}(t)
\widetilde B_{\nu}^{(\alpha)}(0)
\rho_B^{(\alpha)}
\right],
\qquad
\mu,\nu\in\{x,z\},
\label{eq:tempcorr}
\end{eqnarray}
whose diagonal ($\mu=\nu$) and off-diagonal ($\mu\neq \nu$) components characterize,
respectively, the temporal auto- and cross-correlations of the
reservoir fluctuations.
Taking the bath operators to have vanishing thermal averages,
${\rm Tr}_B[
\widetilde B_\mu^{(\alpha)}(t)\rho_B^{(\alpha)}]=0$,
we treat the system--reservoir coupling perturbatively. In the
weak-coupling regime, retaining terms through second order in the
interaction and   employing the Born approximation  consistently to this order \cite{Breuer2002,RivasHuelga2012},  
the second-order time-convolutionless (TCL2) projection-operator
expansion, followed by transformation back to the
Schr\"odinger picture, yields the time-local master equation for the instantaneous
reduced state of the working medium during isochore $\alpha$,
\begin{eqnarray}
\frac{d\rho^{(\alpha)}(t)}{dt}
&=&
-i
\left[
H_S^{(\alpha)},
\rho^{(\alpha)}(t)
\right]
\nonumber\\
&&
-\int_0^t ds
\sum_{\mu,\nu\in\{x,z\}}
\Big\{
\left[
\sigma_\mu,
\widetilde{\sigma}_\nu^{(\alpha)}(-s)
\rho^{(\alpha)}(t)
\right]
C_{\mu\nu}^{(\alpha)}(s)
-
\left[
\sigma_\mu,
\rho^{(\alpha)}(t)
\widetilde{\sigma}_\nu^{(\alpha)}(-s)
\right]
C_{\nu\mu}^{(\alpha)}(-s)
\Big\} \nonumber\\.
\label{eq:rd12}
\end{eqnarray}
A detailed derivation of the TCL2 master equation and the underlying time-convolutionless projection-operator formalism can be found in Refs.~\cite{Shibata1977,BreuerKappler2001,Breuer2002,
ChaturvediShibata1979}.\\

Eq.~\eqref{eq:rd12} is time local in the instantaneous reduced
state, while the finite integration interval 
$0\leqslant s\leqslant t$ retains
finite-time bath-correlation effects within the second-order
perturbative description, without invoking the Markov approximation
\cite{Shibata1977,BreuerKappler2001,Breuer2002}. We further refrain
from making the secular approximation, which would discard terms
oscillating at different Bohr frequencies
\cite{Breuer2002,RivasHuelga2012,FarinaGiovannetti2019}.
The resulting nonsecular dynamics therefore retains the mixed
longitudinal--transverse contributions generated by
$C_{xz}^{(\alpha)}$ and $C_{zx}^{(\alpha)}$, which can couple the
population and coherence dynamics and contribute to both the coherent
and dissipative sectors of the TCL2 generator.
These two contributions can be separated without secularizing the
dynamics \cite{Shibata1977,ChaturvediShibata1979,Breuer2002}.
Defining $\mathcal{R}_{\mu}^{(\alpha)}(t)
\equiv
\sum_{\nu\in\{x,z\}}
\int_{0}^{t} ds\,
C_{\mu\nu}^{(\alpha)}(s)
\widetilde{\sigma}_{\nu}^{(\alpha)}(-s)$,
Eq.~\eqref{eq:rd12} may be written as
\begin{eqnarray}
\frac{d\rho^{(\alpha)}(t)}{dt}
&=&
-i\left[
H_S^{(\alpha)}
+
H_{\rm LS}^{(\alpha)}(t),
\rho^{(\alpha)}(t)
\right]
+
\mathcal{D}_{t}^{(\alpha)}
\left[\rho^{(\alpha)}(t)\right],
\label{eq:TCL2_sep}
\end{eqnarray}
where the finite-time reservoir-induced Hamiltonian renormalization is
\begin{eqnarray}
H_{\rm LS}^{(\alpha)}(t)
&=&
\frac{1}{2i}
\sum_{\mu\in\{x,z\}}
\left[
\sigma_\mu\mathcal{R}_{\mu}^{(\alpha)}(t)
-
\mathcal{R}_{\mu}^{(\alpha)\dagger}(t)\sigma_\mu
\right],
\label{eq:HLS}
\end{eqnarray}
and the complementary dissipative contribution is
\begin{eqnarray}
\mathcal{D}_{t}^{(\alpha)}[\rho]
&=&
\sum_{\mu\in\{x,z\}}
\Bigg[
\mathcal{R}_{\mu}^{(\alpha)}(t)\rho\sigma_\mu
+
\sigma_\mu\rho\,
\mathcal{R}_{\mu}^{(\alpha)\dagger}(t)
-\frac{1}{2}
\left\{
\sigma_\mu\mathcal{R}_{\mu}^{(\alpha)}(t)
+
\mathcal{R}_{\mu}^{(\alpha)\dagger}(t)\sigma_\mu,
\rho
\right\}
\Bigg].
\label{eq:TCL2_diss}
\end{eqnarray}
$H_{\rm LS}^{(\alpha)}(t)$ is obtained from the anti-Hermitian
part of the operator-valued TCL2 kernel after arranging the generator
into adjoint pairs, and represents the finite-time Lamb-shift
renormalization.  
Without secularization, both the coherent and dissipative sectors retain
the mixed longitudinal--transverse contributions generated by the
off-diagonal temporal correlations
$C_{xz}^{(\alpha)}$ and $C_{zx}^{(\alpha)}$; 
consequently,
$\mathcal{D}_{t}^{(\alpha)}$ need not assume GKSL form
\cite{Breuer2002,FarinaGiovannetti2019}.   \\

Accordingly, although the bare system Hamiltonian
$H_S^{(\alpha)}$ is time independent during each fixed-drive isochoric
stroke, the coherent reduced dynamics is governed by
\begin{eqnarray}
H_{\rm eff}^{(\alpha)}(t)
&=&
H_S^{(\alpha)}
+
H_{\rm LS}^{(\alpha)}(t),
\label{eq:Heff}
\end{eqnarray}
which defines the reservoir-renormalized instantaneous energy structure
of the working medium. The complementary dissipative contribution
drives excitation, relaxation, and dephasing, thereby redistributing
populations and modifying coherences. Since both diagonal and
off-diagonal reservoir correlations enter the TCL2 kernel, the
cross-correlated channels can affect both the dissipative dynamics and
the Lamb-shift renormalization. Their cumulative influence is encoded
in the state $\rho_S^{(\alpha)}(\tau_\alpha)$ reached at the end of the
isochore of duration $\tau_\alpha$.  
The reservoir temporal correlation functions entering the TCL2 reduced dynamics
[Eq.~\eqref{eq:rd12}]  can be expressed as 
\begin{eqnarray}
C_{\mu\nu}^{(\alpha)}(t)
&=&  
{\rm Tr}_{B_\alpha}
\left[
B_\mu^{(\alpha)}(t)
B_\nu^{(\alpha)}(0)
\rho_B^{(\alpha)}
\right]
\nonumber\\
&=&
\sum_k
\Big[
g_{k\mu}^{(\alpha)}
\big(g_{k\nu}^{(\alpha)}\big)^*
\big(n_{k,\alpha}+1\big)
e^{-i\omega_k^{(\alpha)}t}
+
\big(g_{k\mu}^{(\alpha)}\big)^*
g_{k\nu}^{(\alpha)}
n_{k,\alpha}
e^{i\omega_k^{(\alpha)}t}
\Big]\,,
\label{eq:cfull1}
\end{eqnarray}
where we used the bosonic commutation relations in
Eq.~\eqref{eq:bosonic_comm} together with the thermal expectation values 
$\av{b_{k,\alpha}^\dagger b_{k,\alpha}}=n_{k,\alpha}$, $\av{b_{k,\alpha} b_{k,\alpha}^\dagger}=n_{k,\alpha}+1$ where
$n_{k,\alpha} \equiv n_\alpha(\omega_k^{(\alpha)}) = (e^{\beta_\alpha\omega_k^{(\alpha)}}-1)^{-1}$ is the
Bose--Einstein occupation number of the $k$-th mode.
  Stationarity of
the reservoir state together with Hermiticity of the bath operators implies
the symmetry $C_{\mu\nu}^{(\alpha)}(t)
=
\big[
C_{\nu\mu}^{(\alpha)}(-t)
\big]^\star$. 
In the frequency domain, the reservoir fluctuations are conveniently encoded
in the matrix-valued spectral density
\begin{eqnarray}
J^{(\alpha)}(\omega)
&=&
\begin{pmatrix}
J_{xx}^{(\alpha)}(\omega)
&
J_{xz}^{(\alpha)}(\omega)
\\
J_{zx}^{(\alpha)}(\omega)
&
J_{zz}^{(\alpha)}(\omega)
\end{pmatrix},
\label{eq:JmatrixAlpha}
\end{eqnarray}
whose elements for $\omega>0$ are defined microscopically as
\cite{BreuerPetruccione2002,Weiss2012,RivasHuelga2012}. 
\begin{eqnarray}
J_{\mu\nu}^{(\alpha)}(\omega)
&=& 
\sum_k
g_{k\mu}^{(\alpha)}
\big(g_{k\nu}^{(\alpha)}\big)^*
\delta
\left(
\omega-\omega_k^{(\alpha)}
\right),
\qquad
\mu,\nu=x,z .
\label{eq:JmunuAlpha}
\end{eqnarray}
The temporal
correlation functions can then be expressed in terms of the corresponding
frequency-resolved spectral-density components as
\begin{eqnarray}
C_{\mu\nu}^{(\alpha)}(t)
&=& 
\int_0^\infty d\omega\,
\Big[
J_{\mu\nu}^{(\alpha)}(\omega)
\big(n_\alpha(\omega)+1\big)
e^{-i\omega t} 
+
\big[J_{\mu\nu}^{(\alpha)}(\omega)\big]^*
n_\alpha(\omega)
e^{i\omega t}
\Big],
\label{eq:CJrelationAlpha}
\end{eqnarray}
where $n_\alpha(\omega)= 
[e^{\beta_\alpha\omega}-1]^{-1}$
is the Bose--Einstein occupation factor~\cite{Breuer2002}.
The diagonal components $J_{xx}^{(\alpha)}(\omega)$ and
$J_{zz}^{(\alpha)}(\omega)$
of the spectral density matrix
are the auto-spectral densities associated with
the transverse and longitudinal coupling channels, respectively. They
characterize the coupling-weighted density of reservoir modes and hence the
fluctuation strength of each individual channel at frequency $\omega>0$.
The off-diagonal components $J_{xz}^{(\alpha)}(\omega)$ and
$J_{zx}^{(\alpha)}(\omega)$ quantify the cross-spectral correlations between
the transverse and longitudinal reservoir fluctuations
~\cite{PazSilva2019,Dutta2026}.\\

By construction \eqref{eq:JmunuAlpha}, the spectral-density matrix
$J^{(\alpha)}(\omega)$ is Hermitian and also
positive semi-definite for all $\omega>0$.
It is therefore fully characterized by its real diagonal components
$J_{xx}^{(\alpha)}(\omega)$ and $J_{zz}^{(\alpha)}(\omega)$ together with the
real and imaginary parts of $J_{xz}^{(\alpha)}(\omega) = [J_{zx}^{(\alpha)}(\omega)]^\star$. 
The positive semidefinite property implies 
that  the diagonal components are non-negative and the
off-diagonal component satisfies the Cauchy--Schwarz bound \cite{Breuer2002}: 
\begin{eqnarray}
J_{xx}^{(\alpha)}(\omega) \geqslant 0\,,\quad
J_{zz}^{(\alpha)}(\omega)  \geqslant 0 \,,\quad
|J_{xz}^{(\alpha)}(\omega)|^2 \leqslant  J_{xx}^{(\alpha)}(\omega) J_{zz}^{(\alpha)}(\omega)\, 
\quad \mbox{(for all $\omega$)}\,.
\label{eq:Jcons}
\end{eqnarray}
Thus, the magnitude of the cross-spectral correlation is bounded by the
auto-spectral strengths of the two coupling channels
~\cite{Breuer2002,PazSilva2019,Dutta2026}.

\subsection{Thermodynamic bookkeeping and cycle energetics} 
\label{sec:bkp}
For the thermodynamic bookkeeping, we adopt the conventional weak-coupling
bare-Hamiltonian prescription \cite{Alicki1979,Talkner2009}, defining the
internal energy of the working medium during isochore $\alpha$ as
\begin{eqnarray}
E_S^{(\alpha)}(t)
&=&
{\rm Tr}_S
\left[
\rho_S^{(\alpha)}(t)
H_S^{(\alpha)}
\right].
\label{eq:internal_energy}
\end{eqnarray}
Within this convention, the reservoir-induced Lamb shift and the
system--reservoir interaction energy are not treated as separate stored-energy
contributions. The energetic cost of switching the coupling on and off is
likewise neglected, consistently with the idealized weak-coupling description
adopted throughout the cycle. \\

The state $\rho_S^{(\alpha)}(\tau_\alpha)$ 
at the end of isochore $\alpha$,
carries the accumulated imprint of
both the Lamb-shift and dissipative dynamics and determines the energy
available to the subsequent unitary stroke. 
Upon decoupling reservoir $\alpha$, the system--reservoir interaction and the associated Lamb-shift contribution are removed, while the reduced state reached at the end of the isochore is retained as the initial state for the subsequent isolated work stroke.
Accordingly,
changes in $E_S$ at fixed $H_S$ during the isochoric strokes are
identified with heat, whereas changes arising from the explicit
time dependence of $H_S$ during the unitary driving strokes are
identified with work. 
Since no work is performed during an isochoric stroke, the heat
$Q_\alpha$ exchanged between reservoir $\alpha$ and the working medium is
given by the corresponding change in the internal energy of the working
medium,
\begin{eqnarray}
Q_\alpha
&=&
{\rm Tr}_S
\left[
\rho_S^{(\alpha)}(\tau_\alpha)
H_S^{(\alpha)}
\right]
-
{\rm Tr}_S
\left[
\rho_S^{(\alpha)}(0)
H_S^{(\alpha)}
\right],\quad \alpha = h,c
\label{eq:Qalpha}
\end{eqnarray}
where $\rho_S^{(\alpha)}(0)$ and
$\rho_S^{(\alpha)}(\tau_\alpha)$ respectively denote the reduced states of the working
medium at the beginning and end of isochore $\alpha$
of duration $\tau_\alpha$.
With this convention, $Q_\alpha>0$ denotes heat absorbed by the working
medium from reservoir $\alpha$, whereas $Q_\alpha<0$ corresponds to heat
released to the reservoir. 
 With the convention that the heat entering the working medium is positive, the heat-engine
regime requires  $Q_h>0$  and  $Q_c<0$.   
 To evaluate $Q_\alpha$, the state
$\rho_S^{(\alpha)}(\tau_\alpha)$ is obtained by solving
Eq.~\eqref{eq:rd12} over $0\leqslant t\leqslant\tau_\alpha$ with the initial
state $\rho_S^{(\alpha)}(0)$.
During the unitary expansion (exp) and compression (comp) strokes, the working medium is
isolated from both reservoirs, so that no heat is exchanged and the change in
its internal energy is entirely identified with work. For a unitary stroke
$\kappa={\rm exp},{\rm comp}$ of duration $\tau_\kappa$, the work performed on
the working medium is therefore
\begin{eqnarray}
W_\kappa
&=&
{\rm Tr}_S
\left[
\rho_S^{(\kappa)}(\tau_\kappa)
H_{S,f}^{(\kappa)}
\right]
-
{\rm Tr}_S
\left[
\rho_S^{(\kappa)}(0)
H_{S,i}^{(\kappa)}
\right],
\label{eq:Wkappa}
\end{eqnarray}
where $H_{S,i}^{(\kappa)}$ and $H_{S,f}^{(\kappa)}$ denote the bare
Hamiltonians at the beginning and end of the corresponding work stroke.
Specifically,
\begin{eqnarray}
H_{S,i}^{({\rm exp})}=
H_S^{(h)}, \quad
H_{S,f}^{({\rm exp})}
=
H_S^{(c)}, \quad
H_{S,i}^{({\rm comp})}
=
H_S^{(c)}, \quad
H_{S,f}^{({\rm comp})}
=
H_S^{(h)}.
\end{eqnarray}
The initial state $\rho_S^{(\kappa)}(0)$ of each unitary stroke of duration $\tau_\kappa$
is the state reached at the end of
the preceding isochore, while the final state  $\rho_S^{(\kappa)}(\tau_\kappa)$ 
is obtained from the corresponding
unitary evolution generated by $H_S(t)$  in Eq.\ \eqref{eq:HSbare}. 
With the convention adopted here,
$W_\kappa>0$ denotes work performed on the working medium, whereas
$W_\kappa<0$ denotes work extracted from it. The net work output per cycle is
therefore
\begin{eqnarray}
W_{\rm cyc}
&=&
-\left(
W_{\rm exp}
+
W_{\rm comp}
\right),
\label{eq:Wcyc}
\end{eqnarray}
with the heat-engine regime requiring $W_{\rm cyc}>0$.
Following the standard
convention for finite-time quantum Otto engines
~\cite{Quan2007,Kosloff2013,Peterson2019,Camati2019},  the corresponding work-to-heat
ratio ($\eta$) 
and average output power are defined respectively 
as
\begin{eqnarray}
\eta
=
\frac{W_{\rm cyc}}{Q_h}\,,\quad
P
=
\frac{W_{\rm cyc}}{\tau_{\rm cyc}},
\label{eq:etapower}
\end{eqnarray}
where  
$\tau_{\rm cyc} = \tau_h + \tau_c + \tau_{\rm exp} + \tau_{\rm comp}$ is the total cycle duration. \\

The physical energetic costs of coupling and
decoupling the reservoir at the beginning and end of isochore are determined by the switching protocol through which the full system--reservoir interaction Hamiltonian $H_{SB}^{(\alpha)}$ is turned on and off. These interface contributions are negligible in the conventional weak-coupling limit but must be accounted for once interaction-energy effects become appreciable, since sudden decoupling can modify both the net work output and the efficiency of a quantum Otto engine.  
The energetic consequences of finite system--reservoir coupling and
decoupling have been analyzed beyond the weak-coupling limit in
Refs.~\cite{Newman2017,PerarnauLlobet2018}.
Since our focus is the thermodynamic role of reservoir cross-channel
correlations, coupling--decoupling interface contributions are neglected.
The cycle energetics are therefore evaluated consistently within the same
weak-coupling bare-Hamiltonian framework.

\section{Bloch Vector Formulation of Isochoric TCL2 Dynamics} 
\label{sec:bloch1} 
To make the influence of the reservoir auto- and cross-spectral terms more
transparent, we recast the isochoric TCL2 dynamics of the reduced
working-medium state in Bloch-vector form. This representation yields a compact
set of equations governing its populations and coherences. 
For the Pauli observables $\sigma_k$ ($k=x,y,z$), the evolution of the expectation values
$\av{\sigma_{k,\alpha}}_t \equiv \trd{S}{[\sigma_k\rho^{(\alpha)}(t)]}$  
follows directly from the TCL2 master equation~\eqref{eq:rd12}, as
\begin{eqnarray}
 \dv{\av{\sigma_{k,\alpha}}_t }{t} 
 &=&
 - i\trd{S}{\Big(\sigma_k\com{H_S^{(\alpha)}}{\rho^{(\alpha)}(t)}\Big)} \nonumber\\
 && -   \int_0^t ds 
\sum_{\mu,\nu \in \{x,z\}} 
\trd{S}{\Big(\sigma_k \com{\sigma_\mu}{ \tilde \sigma_\nu^{(\alpha)}(-s) \rho^{(\alpha)}(t) }\Big)} C_{\mu\nu}^{(\alpha)}(s) \nonumber\\
&& 
\hspace{2cm}-   \trd{S}{\Big(\sigma_k\com{\sigma_\mu}{  \rho^{(\alpha)}(t)\tilde \sigma_\nu^{(\alpha)}(-s)}\Big)}\big[C_{\mu\nu}^{(\alpha)}(s)\big]^\star\,,\quad k = x,y,z 
 \label{eq:rd13} 
\end{eqnarray}
where we have used
$C_{\nu\mu}^{(\alpha)}(-s)
=
[C_{\mu\nu}^{(\alpha)}(s)]^*$.
The three expectation values $\{\av{\sigma_{k,\alpha}}_t\}$
define the components of the Bloch vector $\mathbf{r}_\alpha(t)
=(x_\alpha(t),y_\alpha(t),z_\alpha(t))$
associated with the instantaneous reduced state
\begin{eqnarray}
\rho_S^{(\alpha)}(t)
&=&
\frac{1}{2}
\left[
\mathbb{I}
+
x_\alpha(t)\sigma_x
+
y_\alpha(t)\sigma_y
+
z_\alpha(t)\sigma_z
\right] = \frac12 \left[\mathbb{I} 
+ \mathbf{r}_\alpha(t)
\cdot\boldsymbol{\sigma}\right]\,\,.
\label{eq:Bloch_state}
\end{eqnarray}
where  
$
x_\alpha(t)
=
\langle \sigma_{x,\alpha}\rangle_t\,,\quad
y_\alpha(t)
=
\langle \sigma_{y,\alpha}\rangle_t\,,\quad
z_\alpha(t)
=
\langle \sigma_{z,\alpha}\rangle_t.
$ and $\boldsymbol{\sigma}=(\sigma_x,\sigma_y,\sigma_z)$.
Eq.~\eqref{eq:rd13} therefore describes the evolution of the
Bloch-vector components of the reduced state $\rho_S^{(\alpha)}(t)$
obtained from the TCL2 master equation, Eq.~\eqref{eq:rd12}.
The freely evolved Pauli operators entering Eq.~\eqref{eq:rd13} are given by
\begin{eqnarray}
\widetilde{\sigma}_z^{(\alpha)}(-s)
&=&
e^{i\omega_\alpha s\sigma_z/2}
\sigma_z
e^{-i\omega_\alpha s\sigma_z/2}
=
\sigma_z,
\label{eq:rd14a}
\\
\widetilde{\sigma}_x^{(\alpha)}(-s)
&=&
e^{i\omega_\alpha s\sigma_z/2}
\sigma_x
e^{-i\omega_\alpha s\sigma_z/2}
=
(\cos\omega_\alpha s)\sigma_x
-
(\sin\omega_\alpha s)\sigma_y .
\label{eq:rd14b}
\end{eqnarray}
Substituting these expressions into Eq.~\eqref{eq:rd13}, evaluating the
corresponding system traces, and expressing the reservoir correlation
functions $C_{\mu\nu}^{(\alpha)}(s)$ in terms of the spectral-density
components $J_{\mu\nu}^{(\alpha)}(\omega)$ through
Eq.~\eqref{eq:CJrelationAlpha}, the  
TCL2 dynamics of  Bloch-vector components can be cast in
the affine first-order form  \cite{Dutta2026} 
\begin{eqnarray}
\frac{d}{dt}
\begin{pmatrix}
x_\alpha(t)\\
y_\alpha(t)\\
z_\alpha(t)
\end{pmatrix}
&=&
\mathbf{M}^{(\alpha)}(t)
\begin{pmatrix}
x_\alpha(t)\\
y_\alpha(t)\\
z_\alpha(t)
\end{pmatrix}
+
\mathbf{K}^{(\alpha)}(t),
\label{eq:Bloch_TCL2}
\end{eqnarray}
where $\mathbf{M}(t)$  is a $3\times3$ time-dependent matrix and 
$\mathbf{K}(t)$ is the 3-component drift vector 
with components given by  \cite{Dutta2026}  
{\small
\begin{eqnarray} 
 M_{xx}^{(\alpha)}(t) 
&=& 
-4 \int_{0}^{\infty} d\omega \;
J_{zz}^{(\alpha)}(\omega)\,\coth\!\left(\frac{\beta_\alpha\omega}{2}\right)
\phi_c(\omega,t)  \quad , \quad
M_{xy}^{(\alpha)}(t)  
=
\omega_\alpha \nonumber\\
M_{xz}^{(\alpha)}(t)  
&=&
4 \int_{0}^{\infty} d\omega 
\Big[
\coth\!\left(\frac{\beta_\alpha\omega}{2}\right)\,
\Big(\mathrm{Re}\,J_{xz}^{(\alpha)}(\omega)\Big) 
\phi_{cc}(\omega,t) 
- \Big(\mathrm{Im}\,J_{xz}^{(\alpha)}(\omega)\Big) 
\phi_{cs}(\omega,t)  
\Big] \nonumber\\
M_{yx}^{(\alpha)}(t)  
&=&
-\omega_\alpha -   4
\int_{0}^{\infty} d\omega \;
J_{xx}^{(\alpha)}(\omega)\,\coth\!\left(\frac{\beta_\alpha\omega}{2}\right)
\phi_{ss}(\omega,t) \nonumber\\
 M_{yy}^{(\alpha)}(t)  
&=&
-4 \int_{0}^{\infty} d\omega \coth\left(\frac{\beta_\alpha\omega}{2}\right)
\Bigg[ J_{xx}^{(\alpha)}(\omega)\phi_{cc}(\omega,t) + J_{zz}^{(\alpha)}(\omega) \phi_{c}(\omega,t)\Bigg]  \nonumber   \\
M_{yz}^{(\alpha)}(t)  
&=&
-4\int_{0}^{\infty} d\omega 
\Bigg[\coth\left(\frac{\beta_\alpha\omega}{2}\right)\Big(\mathrm{Re} \ J_{xz}^{(\alpha)}(\omega)\Big) 
\phi_{sc}(\omega,t)
- \Big(\mathrm{Im} \ J_{xz}^{(\alpha)}(\omega)\Big)
\phi_{ss}(\omega,t) 
\Bigg] \nonumber\\
M_{zx}^{(\alpha)}(t)  
&=&
4\int_{0}^{\infty} d\omega 
 \coth\left(\frac{\beta_\alpha\omega}{2}\right)
  \Big[ \big(\mathrm{Re} \ J_{xz}^{(\alpha)}\big)
  \phi_c(\omega,t) +  \big(\mathrm{Im} \ J_{xz}^{(\alpha)}\big)
   \phi_s(\omega,t) \Big] \nonumber\\
M_{zy}^{(\alpha)}(t)  
&=&  
0 \quad , \quad 
M_{zz}^{(\alpha)}(t)  
=  
  -4 
 \int_{0}^{\infty} d\omega J_{xx}^{(\alpha)}(\omega)
\coth\left(\frac{\beta_\alpha\omega}{2}\right) 
\phi_{cc}(\omega,t)
\label{eq:Mfinal}
\end{eqnarray}
}
and,
{\small
\begin{eqnarray}
 K_x^{(\alpha)}(t)  
&=&
-4 
  \int_{0}^{\infty} d\omega \Bigg[ \big(\mathrm{Im} \ J_{xz}(\omega)\big)
\phi_{sc}(\omega,t)   + \big(\mathrm{Re} \ J_{xz}(\omega)\big)
\phi_{ss}(\omega,t)  \Bigg] \nonumber\\
K_y^{(\alpha)}(t)  
&=&
-4 
\int_{0}^{\infty} d\omega \Big[ \big(\mathrm{Im} \ J_{xz}^{(\alpha)}(\omega)\big)\phi_{c}(\omega,t)
  -  \big(\mathrm{Re} \ J_{xz}^{(\alpha)}(\omega)\big)
\phi_{s}(\omega,t)    \Big]\nonumber\\
&&
\qquad - 4  
  \int_{0}^{\infty} d\omega \Big[ \big(\mathrm{Im} \ J_{xz}^{(\alpha)}(\omega)\big)
 \phi_{cc}(\omega,t)   + \big(\mathrm{Re} \ J_{xz}^{(\alpha)}(\omega)\big)
 \phi_{cs}(\omega,t)  \Big] \nonumber\\
K_z^{(\alpha)}(t)   
&=&
 4  \int_{0}^{\infty} d\omega J_{xx}^{(\alpha)}(\omega)
 \phi_{ss}(\omega,t)
 \label{eq:Kfinal}
\end{eqnarray}
} 
where  the functions $\phi_{j}(\omega,t)$'s are given by
\begin{eqnarray}
\phi_c(\omega,t) \equiv \int_0^t ds \cos(\omega s)&,&
\phi_s(\omega,t)  \equiv\int_0^t ds \sin(\omega s)\nonumber\\
\phi_{cc}(\omega,t) \equiv \int_0^t ds\cos(\omega_\alpha s)\cos(\omega s)&,& 
\phi_{cs}(\omega,t) \equiv \int_0^t ds\cos(\omega_\alpha s)\sin(\omega s)\nonumber\\ 
\phi_{sc}(\omega,t) \equiv \int_0^t ds \sin(\omega_\alpha s)\cos(\omega s) &,& 
\phi_{ss}(\omega,t) \equiv \int_0^t ds\sin(\omega_\alpha s)\sin(\omega s)  
\label{eq:phifunc}
\end{eqnarray}
Given the initial reduced state of the working medium at the beginning of
isochore $\alpha$, $\rho_S^{(\alpha)}(0)
=
\frac{1}{2}
\left[
\mathbb{I}
+
\mathbf{r}_\alpha(0)\cdot\boldsymbol{\sigma}
\right]$,
the affine Bloch equation in Eq.~\eqref{eq:Bloch_TCL2} determines the time
evolution of the Bloch vector $\mathbf{r}_\alpha(t)$. Solving this equation
over the interval $0\leqslant t\leqslant \tau_\alpha$ yields the boundary value
$\mathbf{r}_\alpha(\tau_\alpha)$ and hence the reduced state at the end of
the isochoric stroke,
$\rho_S^{(\alpha)}(\tau_\alpha)
=
\frac{1}{2}
\left[
\mathbb{I}
+
\mathbf{r}_\alpha(\tau_\alpha)\cdot\boldsymbol{\sigma}
\right]$.
Thus, the reservoir-induced auto- and cross-spectral contributions entering
$\mathbf{M}^{(\alpha)}(t)$ and $\mathbf{K}^{(\alpha)}(t)$ are accumulated over the interval
$[0,\tau_\alpha]$ 
in the boundary state $\rho_S^{(\alpha)}(\tau_\alpha)$, which serves as the
initial state for the subsequent stroke.

\section{Bloch Vector Framework for Finite-Time Thermodynamic Analysis}
\label{sec:bloch2}
 \begin{figure}[h]
\centering
\includegraphics[width=0.4\linewidth]{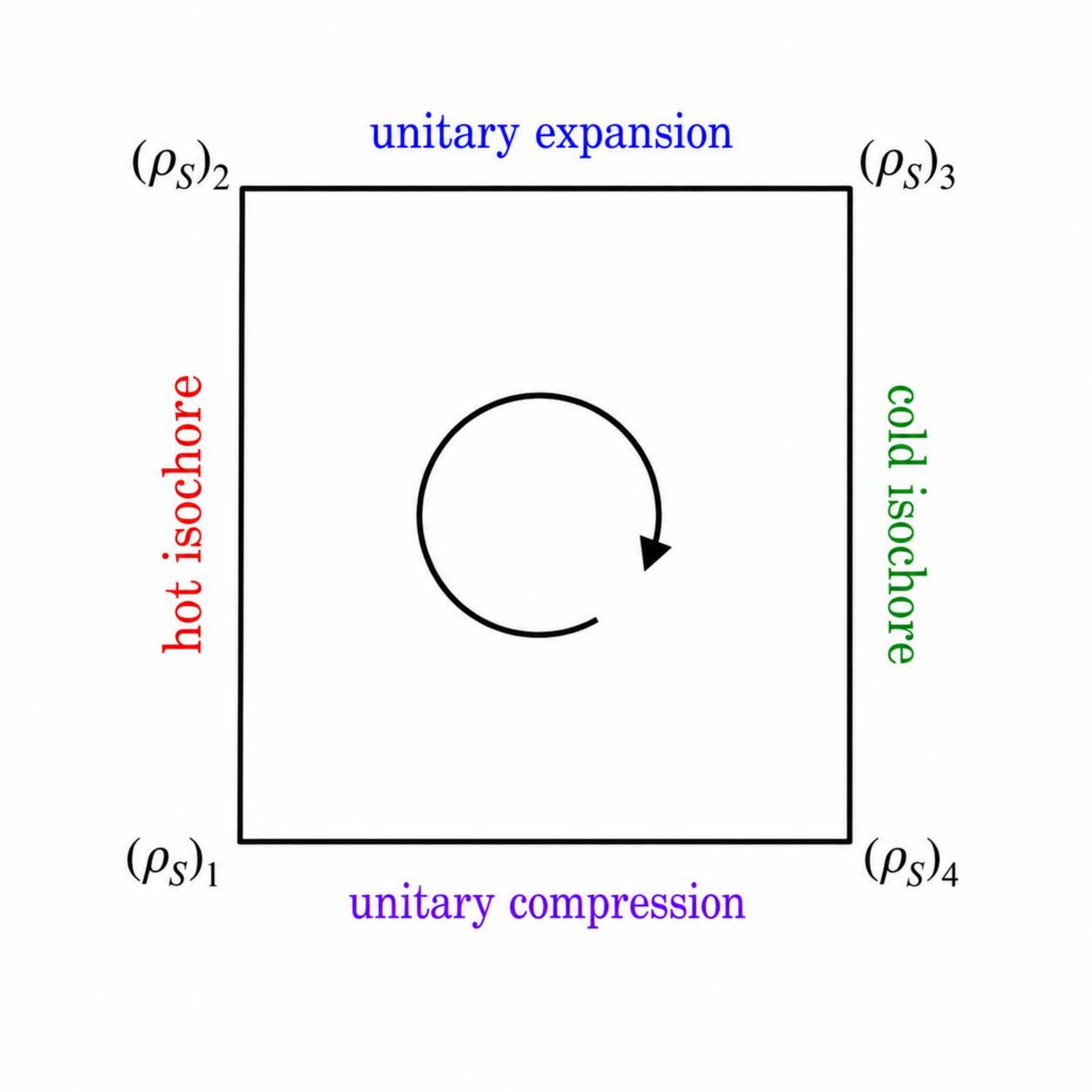}
\caption{Schematic representation of the finite-time quantum Otto cycle, with the working-medium states $(\rho_S)_1$, $(\rho_S)_2$, $(\rho_S)_3$, and $(\rho_S)_4$ defined at the boundaries between successive strokes.}
\label{fig:2}
\end{figure}

To quantify the effect of reservoir cross-spectral correlations on the
finite-time Otto cycle, we propagate the reduced working-medium state
stroke by stroke. The boundary states $(\rho_S)_1$, $(\rho_S)_2$,
$(\rho_S)_3$, and $(\rho_S)_4$, defined in Fig.~\ref{fig:2}, evolve over one
cycle according to
\begin{eqnarray}
(\rho_S)_1
&\xrightarrow{\;h\;}&
(\rho_S)_2
\xrightarrow{\;{\rm exp}\;}
(\rho_S)_3
\xrightarrow{\;c\;}
(\rho_S)_4
\xrightarrow{\;{\rm comp}\;}
(\rho_S)_1'.
\label{eq:stroke_sequence}
\end{eqnarray}
The first cycle starts from $\rho_S(0)\equiv(\rho_S)_1$, while
$\rho_S(\tau_{\rm cyc})\equiv(\rho_S)_1'$ is used as the initial state of the
subsequent cycle, generating the cycle-to-cycle evolution of the engine
\cite{Quan2007,Kosloff2013,Peterson2019}. 
For continuous operation, the four-stroke map is iterated until a periodic
steady state, or limit cycle, is reached \cite{Kosloff2013,Brandner2017}.
Writing $(\rho_S)_1^{(n)}$ for the state at the beginning of the $n$th cycle,
convergence is monitored through the trace-norm
criterion
\begin{eqnarray}
\left\|
(\rho_S)_1^{(n+1)}
-
(\rho_S)_1^{(n)}
\right\|_1
&<&
\epsilon,
\label{eq:limit_cycle_condition}
\end{eqnarray}
where $\epsilon$ is a prescribed numerical tolerance. Once this condition is
satisfied, the same sequence of boundary states is reproduced from cycle to
cycle. 
In constructing this cycle map, we assume that, at the beginning of each
isochore ($\alpha=h,c$), reservoir $\alpha$ is in its thermal state at inverse
temperature $\beta_\alpha$ and initially uncorrelated with the working medium.
The TCL2 evolution retains the finite-time influence of reservoir correlations
during each contact, while system--reservoir correlations generated therein are
discarded upon decoupling. Their dynamical imprint remains encoded in the
emerging reduced state, which serves as the initial state for the subsequent
unitary stroke, thereby closing the cycle map at the level of the working-medium
state.\\

During the hot isochore, the level spacing is fixed at $\omega_h$ and the
state evolves from $(\rho_S)_1$ to $(\rho_S)_2$ under the corresponding TCL2
dynamics. With $0\leqslant s\leqslant\tau_h$ measured from the beginning of
the stroke, the Bloch vector obeys
\begin{eqnarray}
\frac{d{\bf r}_h(s)}{ds}
&=&
{\bf M}^{(h)}(s)\,{\bf r}_h(s)
+
{\bf K}^{(h)}(s),
\label{eq:hot_numerical}
\end{eqnarray}
with ${\bf r}_h(0)={\bf r}_1=(x_1,y_1,z_1)$. Propagation over the
hot-isocore duration gives
${\bf r}_2={\bf r}_h(\tau_h)=(x_2,y_2,z_2)$ and hence
\begin{eqnarray}
(\rho_S)_2
&=&
\frac{1}{2}
\left[
{\mathbb I}
+
{\bf r}_2\cdot{\boldsymbol\sigma}
\right].
\label{eq:rho2}
\end{eqnarray}
The matrix ${\bf M}^{(h)}(s)$ and drift vector ${\bf K}^{(h)}(s)$  are constructed from the spectral-density matrix of the hot reservoir following Eqs.\ \eqref{eq:Mfinal}
and \eqref{eq:Kfinal}.
The subsequent expansion stroke of duration $\tau_{\rm exp}$ is unitary, with the working
medium decoupled from both reservoirs. The state evolves as
\begin{eqnarray}
(\rho_S)_3
&=&
U_{\rm exp}(\rho_S)_2U_{\rm exp}^{\dagger}
=
\frac{1}{2}
\left[
\mathbb{I}
+
{\bf r}_3\cdot{\boldsymbol\sigma}
\right],
\label{eq:rho3}
\end{eqnarray}
where ${\bf r}_3=(x_3,y_3,z_3)$. We consider the linear driving protocol
\begin{eqnarray}
\omega_{\rm exp}(s)
&=&
\omega_h
-
\left(\omega_h-\omega_c\right)
\frac{s}{\tau_{\rm exp}},
\qquad
0\leqslant s\leqslant\tau_{\rm exp}.
\label{eq:linear_expansion}
\end{eqnarray}
Since
$H_S^{\rm exp}(s)=-\omega_{\rm exp}(s)\sigma_z/2$
commutes with itself at different times, the propagator is
\begin{eqnarray}
U_{\rm exp}
&=& 
\exp
\left[
-i\int_{0}^{\tau_{\rm exp}}
H_S^{\rm exp}(s)\,ds
\right] = 
\exp
\left[
\frac{i}{2}\Phi_{\rm exp}\sigma_z
\right], 
\label{eq:Uexp}
\end{eqnarray}
where
$\Phi_{\rm exp}
=
\int_0^{\tau_{\rm exp}}
\omega_{\rm exp}(s) ds$ 
denotes the accumulated dynamical phase over
the expansion stroke.
The unitary stroke therefore leaves the longitudinal Bloch component unchanged
and rotates the transverse components according to
\begin{eqnarray}
x_3
&=&
x_2\cos\Phi_{\rm exp}
+
y_2\sin\Phi_{\rm exp},
\nonumber\\
y_3
&=&
-x_2\sin\Phi_{\rm exp}
+
y_2\cos\Phi_{\rm exp},
\nonumber\\
z_3
&=&
z_2.
\label{eq:Bloch_exp}
\end{eqnarray}
The expansion stroke  thus preserves the instantaneous energy-level
populations and modifies only the phase coherence between the energy
eigenstates through the accumulated phase $\Phi_{\rm exp}$. Nevertheless, the modifications of the reduced state
generated during the preceding hot isochore remain encoded in
$(\rho_S)_2 = (x_2,y_2,z_2)$ and are consequently carried into $(\rho_S)_3= (x_3,y_3,z_3=z_2)$.  
During the cold isochore $0\leqslant s\leqslant \tau_c$, the level spacing is held fixed at
$\omega_c$ and the Bloch-vector obeys  
\begin{eqnarray}
\frac{d{\bf r}_c(s)}{ds}
=
{\bf M}^{(c)}(s)\,{\bf r}_c(s)
+
{\bf K}^{(c)}(s),
\label{eq:cold_numerical}
\end{eqnarray}
with the initial condition ${\bf r}_c(0)
=
{\bf r}_3= (x_3,y_3,z_3=z_2)$.
The matrix ${\bf M}^{(c)}(s)$ and drift vector
${\bf K}^{(c)}(s)$ are constructed from the spectral-density matrix of
the cold reservoir following Eqs.\ \eqref{eq:Mfinal}
and \eqref{eq:Kfinal}.  Integration over the cold-isocore duration gives $  {\bf r}_4 \equiv
{\bf r}_c(\tau_c)  = (x_4,y_4,z_4)$
and yields the state
\begin{eqnarray}
(\rho_S)_4
=
\frac{1}{2}
\left[
{\mathbb I}
+
{\bf r}_4\cdot{\boldsymbol\sigma}
\right].
\label{eq:rho4}
\end{eqnarray}
The final compression stroke is again unitary. Over a duration
$\tau_{\rm comp}$, the level spacing is restored from $\omega_c$ to
$\omega_h$, and the state evolves according to
$(\rho_S)_1'
=
U_{\rm comp}
(\rho_S)_4
U_{\rm comp}^{\dagger}$,
where for a linear compression protocol
$\omega_{\rm comp}(s)
=
\omega_c
+
\left(
\omega_h-\omega_c
\right)
\frac{s}{\tau_{\rm comp}}$ 
  ($0\leqslant s\leqslant\tau_{\rm comp}$),  
  the unitary matrix $U_{\rm comp} =  \exp\big({\frac{i}{2}\Phi_{\rm comp}\sigma_z}\big)$ is determined
  by the dynamical phase
  $\Phi_{\rm comp} = \int_{0}^{\tau_{\rm comp}}
\omega_{\rm comp}(s) ds$, accumulated over the full cold isochore. The Bloch vector $\mathbf{r}_1'=(x_1',y_1',z_1')$ of the resulting state
$(\rho_S)_1'$ is therefore related to $\mathbf{r}_4$ according to
\begin{eqnarray}
x_1'
&=&
x_4\cos\Phi_{\rm comp} 
-
y_4\sin\Phi_{\rm comp},
\nonumber\\
y_1'
&=&
x_4\sin\Phi_{\rm comp} 
+
y_4\cos\Phi_{\rm comp},
\nonumber\\
z_1'
&=&
z_4.
\label{eq:Bloch_comp}
\end{eqnarray}
Thus, the compression stroke leaves the populations in the energy eigenbasis
unchanged, while rotating the transverse Bloch-vector components in the
$x$--$y$ plane by the accumulated phase $\Phi_{\rm comp}$.
For a cycle that has reached the limit cycle,
$(\rho_S)_1'\simeq(\rho_S)_1$, implying in particular that
$z_1'\simeq z_1$.
The total cycle duration is $\tau_{\rm cyc}
=
\tau_h
+
\tau_{\rm exp}
+
\tau_c
+
\tau_{\rm comp}$.\\

Using the weak-coupling thermodynamic definitions of
Eqs.~\eqref{eq:Qalpha}, \eqref{eq:Wkappa}, \eqref{eq:Wcyc}, and
\eqref{eq:etapower}, together with the components
of the Bloch-vectors ($\mathbf{r}_i,\mathbf{r}_1'$) at the stroke
boundaries, the heat exchanged during the hot and cold isochores is
\begin{eqnarray}
Q_h
&=&
\frac{\omega_h}{2}(z_1-z_2),
\qquad
Q_c
=
-\frac{\omega_c}{2}(z_4-z_3).
\label{eq:Qhc}
\end{eqnarray}
 The finite-time
reservoir dynamics encoded in the TCL2 generator determines the
boundary populations $z_2$ and $z_4$ and therefore directly affects
the heat exchanged during the isochoric strokes. In particular, the
cross-spectral density $J_{xz}^{(\alpha)}(\omega)$ modifies these
boundary states through its contributions to both the dissipative and
reservoir-induced coherent parts of the reduced dynamics.
The work performed on the working medium during the expansion and compression
strokes is
\begin{eqnarray}
W_{\rm exp}
&=&
\frac{\omega_h-\omega_c}{2}z_2,
\qquad
W_{\rm comp}
=
-\frac{\omega_h-\omega_c}{2}z_4,
\label{eq:Wexpcomp}
\end{eqnarray}
where $z_3=z_2$ and $z_1'=z_4$ follow from
Eqs.~\eqref{eq:Bloch_exp} and \eqref{eq:Bloch_comp}, respectively. The net
work output per cycle is therefore
\begin{eqnarray}
W_{\rm cyc}
&=&
-\left(W_{\rm exp}+W_{\rm comp}\right)
=
\frac{\omega_h-\omega_c}{2}(z_4-z_2).
\label{eq:Wcyc_bloch}
\end{eqnarray}
Accordingly, the work-to-heat ratio and average output power take the forms
\begin{eqnarray}
\eta
&=&
\frac{W_{\rm cyc}}{Q_h}
=
\left(1-\frac{\omega_c}{\omega_h}\right)
\frac{z_4-z_2}{z_1-z_2},
\nonumber\\
P
&=&
\frac{W_{\rm cyc}}{\tau_{\rm cyc}}
=
\frac{\omega_h-\omega_c}{2\tau_{\rm cyc}}
(z_4-z_2).
\label{eq:etapower_bloch}
\end{eqnarray}

Before the limit cycle is reached, the state entering successive cycles is
generally different, and the thermodynamic quantities therefore retain an
explicit cycle-number dependence. Denoting the stroke-boundary Bloch
components in the $n$th cycle by $z_i^{(n)}$, we may define corresponding cycle-number-dependent quantities
\begin{eqnarray}
\eta_{\rm tr}^{(n)}
&=&
\frac{W_{\rm cyc}^{(n)}}{Q_h^{(n)}},
\qquad
P^{(n)}
=
\frac{W_{\rm cyc}^{(n)}}{\tau_{\rm cyc}}.
\label{eq:transient_performance}
\end{eqnarray}
Since $(\rho_S)_1^{(n+1)}\neq(\rho_S)_1^{(n)}$ in this regime, a single
four-stroke traversal does not yet constitute a closed thermodynamic cycle.
The corresponding first-law balance is
$E_1^{(n+1)}-E_1^{(n)}
=
Q_h^{(n)}
+
Q_c^{(n)}
-
W_{\rm cyc}^{(n)}$.
Accordingly, $\eta_{\rm tr}^{(n)}$ is a cycle-resolved transient
work-to-heat ratio rather than the efficiency of a closed cycle, since part
of the exchanged energy may contribute to a change in the internal energy of
the working medium.\\

Cross-spectral correlations can strongly affect this transient evolution.
Through the $J_{xz}^{(\alpha)}(\omega)$ contributions to the TCL2 generator,
they modify both dissipative and reservoir-induced coherent dynamics during
the isochores and thereby alter the stroke-boundary components
$z_1^{(n)}$, $z_2^{(n)}$, and $z_4^{(n)}$. Consequently, the heat, work,
transient work-to-heat ratio, and power can depend on the strength, phase,
and characteristic frequency scale of the cross-spectral correlations as the
engine approaches periodic operation.\\

At the limit cycle (LC),
$(\rho_S)_1^{(n+1)}\simeq(\rho_S)_1^{(n)}$, so that the internal-energy
change over one cycle vanishes and
$Q_h^{\rm LC}
+
Q_c^{\rm LC}
-
W_{\rm cyc}^{\rm LC}
=
0$.
Moreover, $z_1'=z_4$ from the compression stroke, while periodicity gives
$z_1'=z_1$, and hence $z_4^{\rm LC}=z_1^{\rm LC}$. Substitution into the
work-to-heat ratio yields the asymptotic efficiency
\begin{eqnarray}
\eta_{\rm LC}
&=&
1-\frac{\omega_c}{\omega_h},
\label{eq:eta_LC}
\end{eqnarray}
which is  is the Otto efficiency for the present frictionless unitary driving protocol.
Thus, within the thermodynamic convention adopted
here, the cross-spectral correlations do not alter the limit-cycle
efficiency itself. They can nevertheless modify the heat exchanged
per cycle and the extracted work through their influence on the
limit-cycle populations, and hence can modify the asymptotic power,
\begin{eqnarray}
P_{\rm LC}
&=&
\frac{\omega_h-\omega_c}{2\tau_{\rm cyc}}
\left(
z_4^{\rm LC}-z_2^{\rm LC}
\right).
\label{eq:power_LC}
\end{eqnarray}
They can also affect the number of cycles required to reach the
periodic steady state. The transient and limit-cycle regimes should
therefore be analyzed separately: cross-spectral correlations may
produce cycle-dependent modifications of both the transient
work-to-heat ratio and power before periodic operation is established,
whereas in the asymptotic regime their influence is manifested
primarily through the heat currents, work output, and power rather
than through the Otto efficiency itself.

\section{Numerical results and discussions} 
\label{sec:results}
In this section, we present the numerical results and examine how longitudinal--transverse cross-spectral reservoir correlations affect the finite-time performance of the quantum Otto engine.

\paragraph{Auto- and Cross-Spectral Density Parameterization:}
For the present investigation, we model the diagonal (auto-)spectral densities by
the Ohmic-type form
\begin{eqnarray}
J_{\mu\mu}^{(\alpha)}(\omega)
&=&
\eta_{\mu}^{(\alpha)}
\left[\omega_{c\mu}^{(\alpha)}\right]^{1-s_{\mu}^{(\alpha)}}
\omega^{s_{\mu}^{(\alpha)}}
\exp\left[-\frac{\omega}{\omega_{c\mu}^{(\alpha)}}\right],
\qquad
\mu=x,z,
\label{eq:Jdiag}
\end{eqnarray}
where $\eta_{\mu}^{(\alpha)}$ denotes the coupling strength,
$\omega_{c\mu}^{(\alpha)}$ is the cutoff frequency, and
$s_{\mu}^{(\alpha)}$ determines the low-frequency scaling. The regimes
$0<s_{\mu}^{(\alpha)}<1$, $s_{\mu}^{(\alpha)}=1$, and
$s_{\mu}^{(\alpha)}>1$ correspond to sub-Ohmic, Ohmic, and super-Ohmic
spectra, respectively~\cite{Leggett1987,Weiss2012,Breuer2002}.
The exponential cutoff accounts phenomenologically for the finite spectral
bandwidth of the reservoir and regularizes the high-frequency contribution.
Such spectral forms are widely employed to model broadband bosonic
environments in quantum-optical and condensed-matter settings
~\cite{Caldeira1983,Clerk2010}. 
The cross-spectral density is parametrized as
\begin{eqnarray}
J_{xz}^{(\alpha)}(\omega)
&=&
r_{\alpha}
\exp\left(-\frac{\omega}{\omega_{\rm corr}^{(\alpha)}}\right)
e^{i\phi_{\alpha}}
\sqrt{
J_{xx}^{(\alpha)}(\omega)
J_{zz}^{(\alpha)}(\omega)
}\,, \mbox{ with }
J_{zx}^{(\alpha)}(\omega)=
\left[J_{xz}^{(\alpha)}(\omega)\right]^\star .
\label{eq:Jcross}
\end{eqnarray}
For $0\leqslant r_{\alpha}\leqslant 1$, this parametrization automatically satisfies
the Cauchy--Schwarz bound in Eq.~\eqref{eq:Jcons}. The parameter
$r_{\alpha}$ controls the strength of the cross-channel correlations, with
$r_{\alpha}=0$ corresponding to vanishing cross correlations,
$\phi_{\alpha}$ specifies their relative phase, and
$\omega_{\rm corr}^{(\alpha)}$ sets the characteristic frequency range over
which the cross-spectral contribution remains appreciable.\\

The finite correlation time of the hot and cold reservoirs gives rise to memory effects encoded in their frequency-dependent auto- and cross-spectral densities. These spectral properties determine the reservoir correlation functions entering the reduced dynamics during the isochoric strokes and thereby modify the cycle efficiency, output power, work extraction, and ergotropy. In particular, the correlation amplitudes $r_{\alpha}$, phases $\phi_{\alpha}$, spectral profiles, and characteristic frequency scales $\omega_{\rm corr}^{(\alpha)}$,
serve as reservoir-engineering control parameters 
in the investigation of thermodynamic consequences of cross-channel correlations.

\paragraph{Units and numerical conventions:}
We work in natural units, $\hbar=k_{\rm B}=1$, and choose the
cold-isochore level spacing $\omega_c$ as the reference energy scale,
setting $\omega_c=1$. All other frequencies and energy scales,
including $\omega_h$, the auto-spectral cutoff frequencies
$\omega_{cx}^{(\alpha)}$ and $\omega_{cz}^{(\alpha)}$, and the
cross-correlation frequency $\omega_{\rm corr}^{(\alpha)}$, are
therefore expressed in units of $\omega_c$, while all time scales are
given in units of $\omega_c^{-1}$. The coupling strengths, Ohmicity
exponents, cross-correlation amplitudes, and cross-spectral phases are
dimensionless, with phases measured in radians. The inverse
temperatures $\beta_\alpha$ are expressed in units of
$(\hbar\omega_c)^{-1}$; thus, for example, $\beta_\alpha=40$
corresponds to $k_{\rm B}T_\alpha/(\hbar\omega_c)=0.025$. All parameters quoted below are understood in these
scaled units.

\paragraph{Fixed parameters and numerical protocol:}
Unless stated otherwise, the numerical results are obtained for fixed
engine and auto-spectral parameters, so that changes in the
thermodynamic observables can be attributed directly to the
cross-spectral structure of the reservoirs. We set
\begin{eqnarray}
\omega_h=5,\qquad
\omega_c=1,\qquad
\beta_h=1,\qquad
\beta_c=40,
\label{eq:fixed_engine_par1}
\end{eqnarray}
and consider the two cycle protocols
\begin{eqnarray}
(\tau_h,\tau_c,\tau_{\rm exp},\tau_{\rm comp},\tau_{\rm cyc})
&=&
(2,2,2,2,8),
\label{eq:fixed_engine_par2a}\\
(\tau_h,\tau_c,\tau_{\rm exp},\tau_{\rm comp},\tau_{\rm cyc})
&=&
(5,5,5,5,20).
\label{eq:fixed_engine_par2b}
\end{eqnarray}
For both hot and cold reservoirs, the auto-spectral parameters are fixed as
\begin{eqnarray}
\eta_x^{(h)}
=
\eta_z^{(h)}
=
\eta_x^{(c)}
=
\eta_z^{(c)}
&=&
0.1,
\nonumber\\
s_x^{(h)}
=
s_z^{(h)}
=
s_x^{(c)}
=
s_z^{(c)}
&=&
1,
\nonumber\\
\omega_{cx}^{(h)}
=
\omega_{cz}^{(h)}
=
\omega_{cx}^{(c)}
=
\omega_{cz}^{(c)}
&=&
5.
\label{eq:fixed_spectral_par}
\end{eqnarray}
The cross-spectral parameters
$r_\alpha$, $\omega_{\rm corr}^{(\alpha)}$, and $\phi_\alpha$
are varied as specified in the individual figures, while all remaining
cross-spectral parameters are held fixed at the corresponding benchmark
values.

\paragraph{Phase-sensitive power response to cold-reservoir cross-spectral cut-off frequency $\omega_{\rm corr}^{(c)}$:}
\begin{figure}[h]
\centering
\includegraphics[width=0.8\linewidth]
{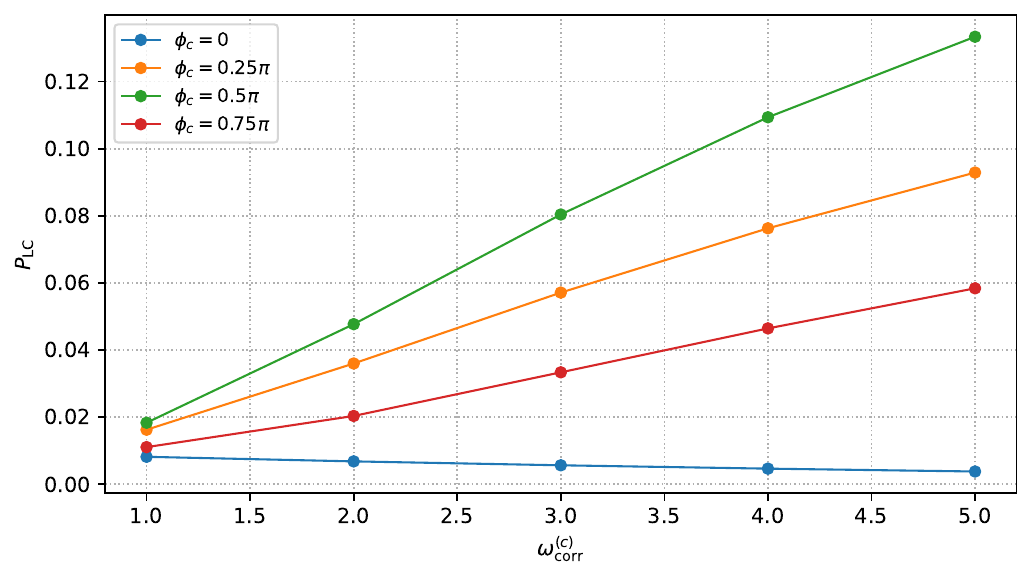}
\caption{
Power generated in the last cycle as a function of the characteristic
cross-correlation frequency $\omega_{\rm corr}^{(c)}$ of the cold
reservoir for different values of the cross-spectral phase
$\phi_c=0,\pi/4,\pi/2$ and $3\pi/4$ 
with fixed correlation strength $r_c=1$. 
 Corresponding hot-reservoir parameters
are fixed at $r_h=1$, $\omega_{\rm corr}^{(h)}=5$, $\phi_h=\pi/2$. The cycle duration is 
$\tau_{\rm cyc} = 8$ with $ \tau_h=\tau_c=\tau_{\rm exp}=\tau_{\rm comp}=2$.
All remaining are parameters fixed as specified in the text.}
\label{fig:power_wcorr_c}
\end{figure}
We first examine the asymptotic-cycle power $P_{\rm LC}$ and its dependence on the characteristic cross-spectral frequency scale $\omega_{\rm corr}^{(\alpha)}$ and phase $\phi_\alpha$ introduced in Eq.~\eqref{eq:Jcross}. Figure~\ref{fig:power_wcorr_c} shows $P_{\rm LC}$ as a function of the cold-reservoir cutoff frequency $\omega_{\rm corr}^{(c)}$ for several values of $\phi_c$, with $r_c=1$, while the hot-reservoir parameters are fixed at $r_h=1$, $\omega_{\rm corr}^{(h)}=5$, and $\phi_h=\pi/2$. 
The power response is strongly phase dependent. Increasing $\omega_{\rm corr}^{(c)}$ produces the largest enhancement near $\phi_c=\pi/2$, whereas for $\phi_c=0$ the power decreases weakly. Since the auto-spectral densities are unchanged, these differences arise entirely from the off-diagonal cross-spectral terms, which couple relaxation and dephasing during the cold isochore and modify both the dissipative kernel and the Lamb-shift contribution. The cross-spectral phase therefore controls how the two coupling channels combine in the reduced dynamics and, consequently, the power generated over repeated cycles. 
A further feature is the crossover of the phase-resolved curves at small $\omega_{\rm corr}^{(c)}$: the $\phi_c=\pi/2$ configuration yields the lowest power near $\omega_{\rm corr}^{(c)}\simeq1$, but becomes the most favorable at larger cutoff frequencies. This shows that the thermodynamic effect of the cross-spectral correlations is determined jointly by their characteristic frequency scale and phase.

\paragraph{Cycle-to-cycle convergence of the efficiency:} 
\begin{figure}[!h]
\centering
\begin{subfigure}[t]{0.48\columnwidth}
\centering
\includegraphics[width=\linewidth]
{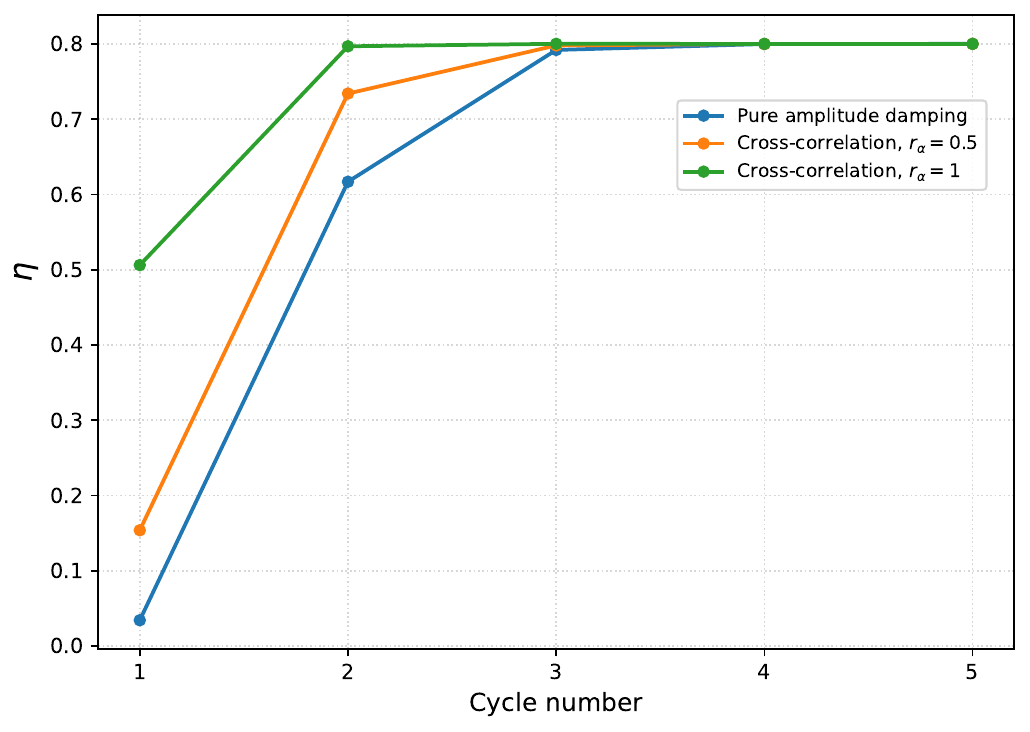}
\caption{$\tau_{\rm cyc}=8$ with
$\tau_h=\tau_c=\tau_{\rm exp}=\tau_{\rm comp}=2$.}
\label{fig:f4a}
\end{subfigure}
\hfill
\begin{subfigure}[t]{0.48\columnwidth}
\centering
\includegraphics[width=\linewidth]
{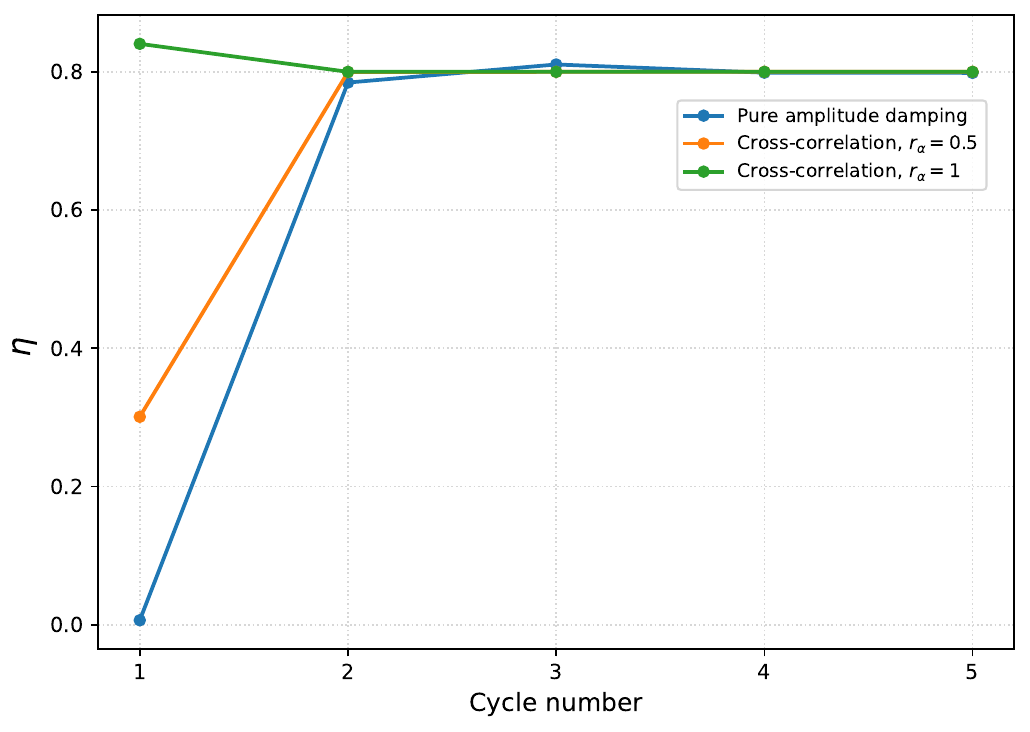}
\caption{$\tau_{\rm cyc}=20$ with
$\tau_h=\tau_c=\tau_{\rm exp}=\tau_{\rm comp}=5$.}
\label{fig:f4b}
\end{subfigure}
\caption{
Cycle-resolved work to heat ratio $\eta$ for pure-dephasing and
pure-amplitude-damping reservoirs, compared with reservoirs possessing
longitudinal--transverse cross-spectral correlations of strengths
$r_c=r_h=0.5$ and $1$. The reservoir parameters are
$\beta_h=1$, $\beta_c=40$, $\omega_{\rm corr}=5$, and
$\phi=\pi/2$, while the two panels correspond to different durations
of the four Otto strokes: (a) $\tau_{\rm cyc}=8$ with
$\tau_h=\tau_c=\tau_{\rm exp}=\tau_{\rm comp}=2$  and
(b) $\tau_{\rm cyc}=20$ with
$\tau_h=\tau_c=\tau_{\rm exp}=\tau_{\rm comp}=5$. All remaining parameters are fixed as
specified in the text.
}
\label{fig:f4}
\end{figure}
Figure~\ref{fig:f4} compares the cycle-resolved efficiency in the
pure-amplitude-damping limit with that obtained in the presence of
longitudinal--transverse cross-spectral correlations. For the shorter
cycle, $\tau_{\rm cyc}=8$ [Fig.~\ref{fig:f4a}], the efficiency exhibits
a pronounced transient dependence on the reservoir correlations. In the
pure-amplitude-damping case, the first-cycle efficiency is small and
approaches the asymptotic value $\eta_{\rm LC}$ [Eq.~\eqref{eq:eta_LC}]
only after several repetitions. Cross-spectral correlations accelerate
this convergence: for $r_\alpha=0.5$ the transient is reduced, while for
$r_\alpha=1$ the efficiency is already substantial in the first cycle
and approaches $\eta_{\rm LC}$ by the second cycle.
For the longer cycle, $\tau_{\rm cyc}=20$
[Fig.~\ref{fig:f4b}], all coupling configurations converge to
$\eta_{\rm LC}$ within approximately two cycles. The maximally
correlated case, $r_\alpha=1$, shows a small first-cycle overshoot,
whereas the pure-amplitude-damping and $r_\alpha=0.5$ cases approach the
asymptotic value from below. Thus, increasing the reservoir-contact
times suppresses the transient sensitivity to the initial state and
drives the engine more rapidly toward repetitive cyclic operation.\\

These results show that, in the parameter regime considered, the main effect of the cross-spectral correlations is to modify the transient approach to cyclic operation rather than the asymptotic efficiency itself. Despite pronounced differences during the first few cycles, all coupling configurations converge to the same limiting value $
\eta_{\rm LC} = 1-\frac{\omega_c}{\omega_h}$.
The cross-spectral terms alter the coupled relaxation--dephasing dynamics during the isochores and hence the states entering the subsequent work strokes, thereby changing the rate of cycle-to-cycle convergence. The comparison of Figs.~\ref{fig:f4a} and \ref{fig:f4b} further shows that this transient correlation dependence is controlled by the finite duration of the individual strokes.

\paragraph{Correlation-enhanced power and finite-cycle-time effects:} 
\begin{figure}[!h]
\centering
\begin{subfigure}[t]{0.48\columnwidth}
\centering
\includegraphics[width=\linewidth]
{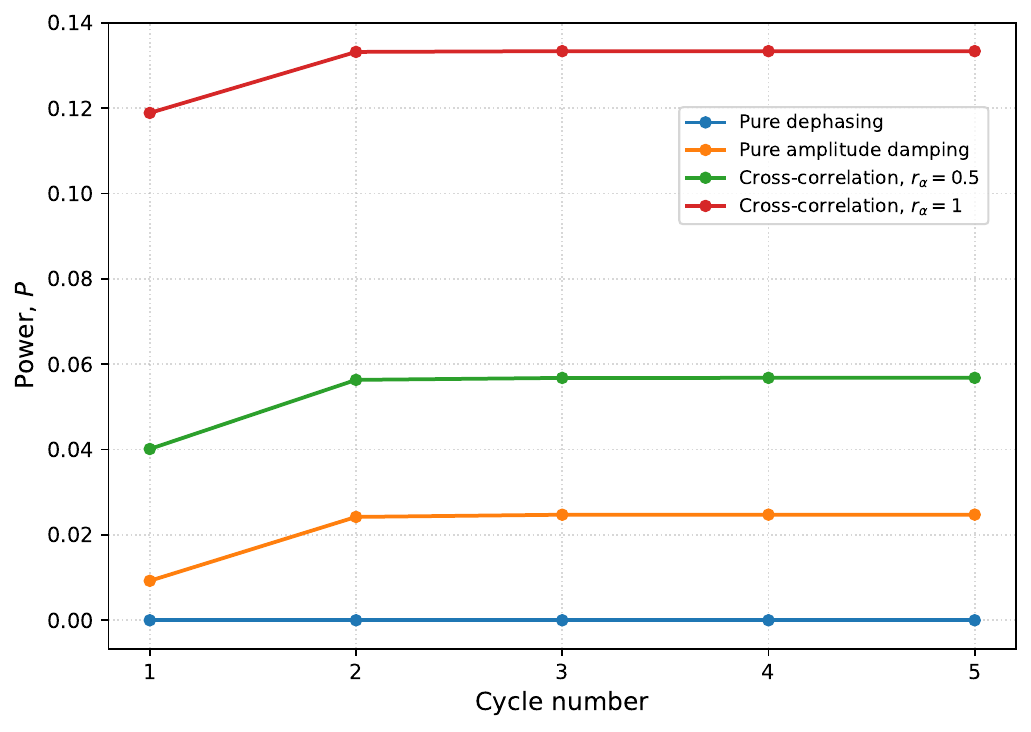}
\caption{$\tau_{\rm cyc}=8$, with
$\tau_h=\tau_c=\tau_{\rm exp}=\tau_{\rm comp}=2$.}
\label{fig:f5a}
\end{subfigure}
\hfill
\begin{subfigure}[t]{0.48\columnwidth}
\centering
\includegraphics[width=\linewidth]
{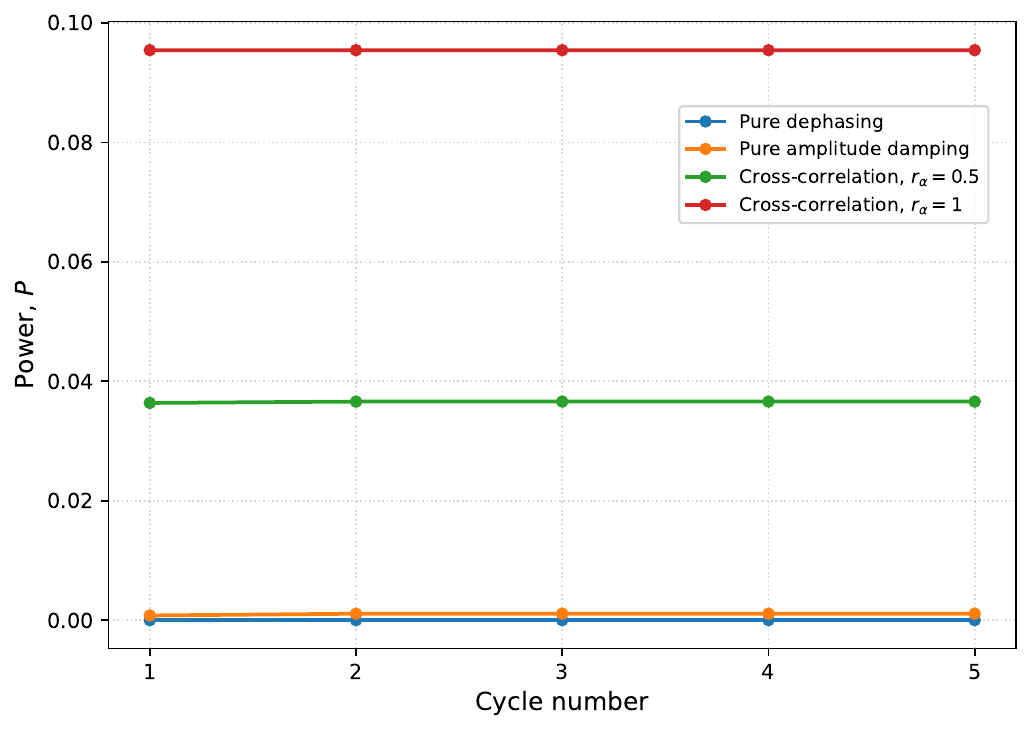}
\caption{$\tau_{\rm cyc}=20$, with
$\tau_h=\tau_c=\tau_{\rm exp}=\tau_{\rm comp}=5$.}
\label{fig:f5b}
\end{subfigure}
\caption{
Cycle-resolved power $P$ for the pure-dephasing and
pure-amplitude-damping limits and in the presence of
longitudinal--transverse cross-spectral correlations with
$r_\alpha=0.5$ and $1$. The reservoir cross-spectral parameters are 
$\omega_{\rm corr}^{(\alpha)}=5$, and
$\phi_\alpha=\pi/2$. Panels (a) and (b) correspond,
respectively, to cycle durations $\tau_{\rm cyc}=8$
with $\tau_h=\tau_c=\tau_{\rm exp}=\tau_{\rm comp}=2$ and $20$ with $\tau_h=\tau_c=\tau_{\rm exp}=\tau_{\rm comp}=5$.
All remaining parameters are fixed as specified in the text.
}
\label{fig:f5}
\end{figure}
Fig.~\ref{fig:f5} shows the cycle-resolved power for the same
reservoir configurations considered in Fig.~\ref{fig:f4}. In the
pure-dephasing limit, the power vanishes because the reservoir does
not modify the energy-basis populations and therefore cannot mediate
the heat exchange required for work extraction. Pure amplitude
damping  produces a finite but comparatively small
power. \\

The inclusion of longitudinal--transverse cross-spectral correlations
enhances the generated power, with the enhancement increasing with the
correlation strength. For the shorter cycle,
$\tau_{\rm cyc}=8$ [Fig.~\ref{fig:f5a}], the power exhibits a brief
cycle-to-cycle transient before approaching a stationary value. This
asymptotic power remains strongly correlation dependent, increasing
from the pure-amplitude-damping case to $r_\alpha=0.5$ and further to
$r_\alpha=1$. In contrast to the efficiency in Fig.~\ref{fig:f4},
which converges to nearly the same limiting value for the different
dissipative configurations, the power retains a pronounced dependence
on the cross-spectral correlation strength.
For the longer cycle, $\tau_{\rm cyc}=20$
[Fig.~\ref{fig:f5b}], the transient feature is strongly suppressed and the
power is nearly cycle independent from the outset. The ordering of the
different reservoir configurations, however, remains unchanged, with
the largest power obtained for $r_\alpha=1$, followed by
$r_\alpha=0.5$ and the pure-amplitude-damping limit. Thus, the
cross-spectral correlations affect both the approach to cyclic
operation and the asymptotic work delivered per unit time.\\

Comparison of Figs.~\ref{fig:f5a} and \ref{fig:f5b} shows that a longer
cycle duration reduces the power for a given reservoir configuration.
Although extended strokes allow more complete reservoir-induced
evolution, the extracted work is delivered over a larger
$\tau_{\rm cyc}$, thereby suppressing
$P=-W_{\rm cyc}/\tau_{\rm cyc}$. This reflects the finite-time
trade-off between establishing cyclic operation and maintaining a large
work output per unit time. Within this trade-off, the
longitudinal--transverse cross-spectral correlations provide a robust
enhancement of the engine power.

\paragraph{Temperature dependence of the asymptotic power:}
\begin{figure}[h]
\centering
\begin{subfigure}[t]{0.48\columnwidth}
\centering
\includegraphics[width=\linewidth]
{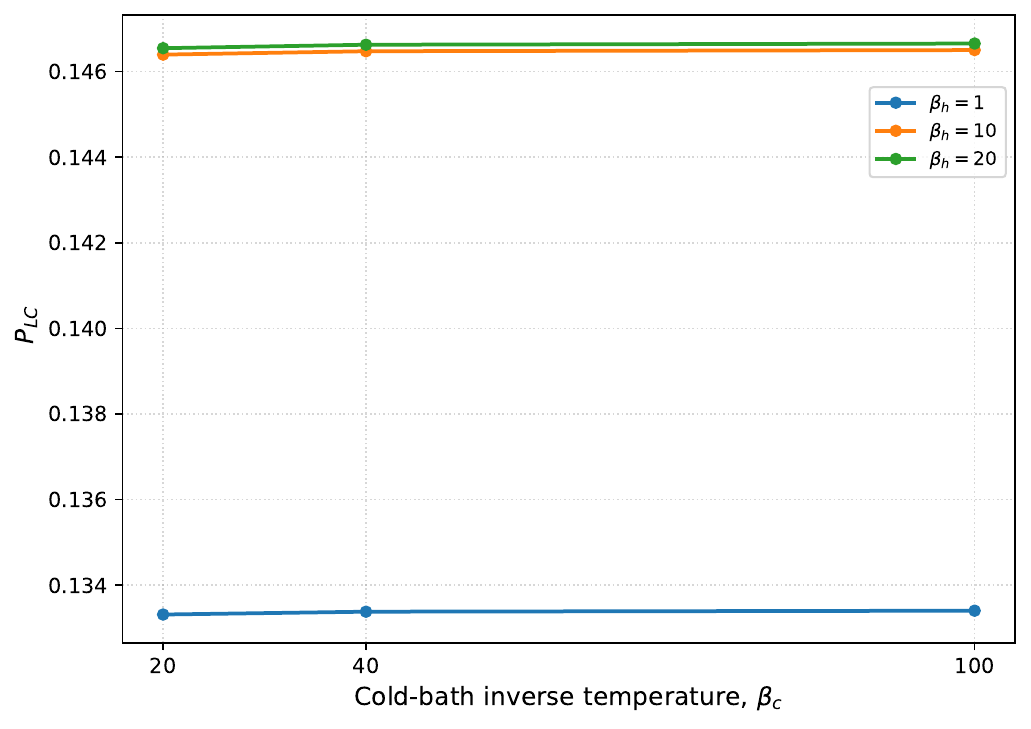}
\caption{$\tau_{\rm cyc}=8$, with
$\tau_h=\tau_c=\tau_{\rm exp}=\tau_{\rm comp}=2$.}
\label{fig:f6a}
\end{subfigure}
\hfill
\begin{subfigure}[t]{0.48\columnwidth}
\centering
\includegraphics[width=\linewidth]
{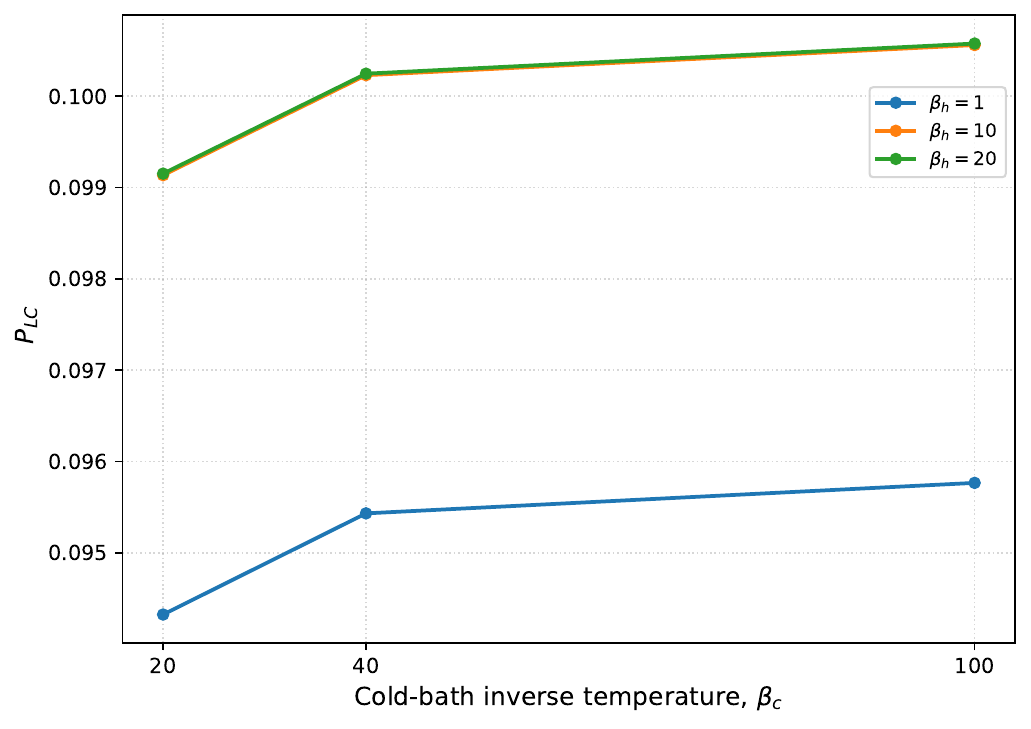}
\caption{$\tau_{\rm cyc}=20$, with
$\tau_h=\tau_c=\tau_{\rm exp}=\tau_{\rm comp}=5$.}
\label{fig:f6b}
\end{subfigure}
\caption{
Asymptotic-cycle power $P_{\rm LC}$ as a function of the
cold-reservoir inverse temperature $\beta_c$ for
$\beta_h=1$, $10$, and $20$. The cross-spectral parameters are
fixed at $r_\alpha=1$,
$\omega_{\rm corr}^{(\alpha)}=5$, and
$\phi_\alpha=\pi/2$ ($\alpha=h,c$).
Panels (a) and (b) correspond to cycle durations
$\tau_{\rm cyc}=8$ and $20$, respectively.
All remaining parameters are fixed as specified in the text.
}
\label{fig:f6}
\end{figure}
Figure~\ref{fig:f6} shows the asymptotic-cycle power $P_{\rm LC}$ as a function of the reservoir temperatures for fixed cross-spectral parameters. In both panels, increasing $\beta_c$ (decreasing $T_c$) leads to a modest increase in $P_{\rm LC}$, with the strongest variation occurring between $\beta_c=20$ and $40$. For larger $\beta_c$, the power tends to saturate, consistent with the cold reservoir entering a low-temperature regime in which further cooling produces only small changes in the population distribution established during the cold isochore.\\

The asymptotic power also depends on the hot-reservoir temperature. For both cycle durations, the $\beta_h=1$ branch lies well below those for $\beta_h=10$ and $20$, while the latter nearly overlap, indicating a weak dependence on $\beta_h$ once the hot reservoir enters this regime. Together with the saturation observed at large $\beta_c$, this shows that $P_{\rm LC}$ is governed not simply by the temperature bias, but by its interplay with finite-time reservoir dynamics and the fixed cross-spectral correlations.\\

Comparison of Figs.~\ref{fig:f6a} and \ref{fig:f6b} shows that increasing the cycle duration from $\tau_{\rm cyc}=8$ to $20$ suppresses the power for all temperature combinations. Although longer strokes permit more extensive reservoir-induced evolution, the extracted work is distributed over a larger cycle time, so that the increase in energy exchange does not compensate for the $1/\tau_{\rm cyc}$ scaling of $P_{\rm LC}=-W_{\rm LC}/\tau_{\rm cyc}$. The stronger $\beta_c$ dependence for the longer cycle further reflects the increased sensitivity of the working-medium state to the cold-reservoir temperature during the extended isochoric contact.

\paragraph{Population dynamics underlying the thermodynamic response:}
The preceding results quantify the effect of reservoir cross-spectral
correlations through cycle-integrated observables such as work, power,
and efficiency. To expose the underlying dynamics, we examine the
population imbalance $\langle\sigma_z\rangle$, which for
$H_S(t)=-\omega(t)\sigma_z/2$ directly determines the bare internal
energy $E_S(t)=-\frac{\omega(t)}{2}\langle\sigma_z(t)\rangle $. 
Its variation during the isochores therefore tracks reservoir-induced
population transfer, while $[H_S(t),\sigma_z]=0$ ensures that it remains
unchanged during the unitary strokes. The evolution of
$\langle\sigma_z\rangle$ thus provides a direct dynamical probe of the
cross-correlation-induced modification of heat exchange over successive
cycles. \\

\begin{figure}[t]
\centering
\includegraphics[width=0.75\columnwidth]
{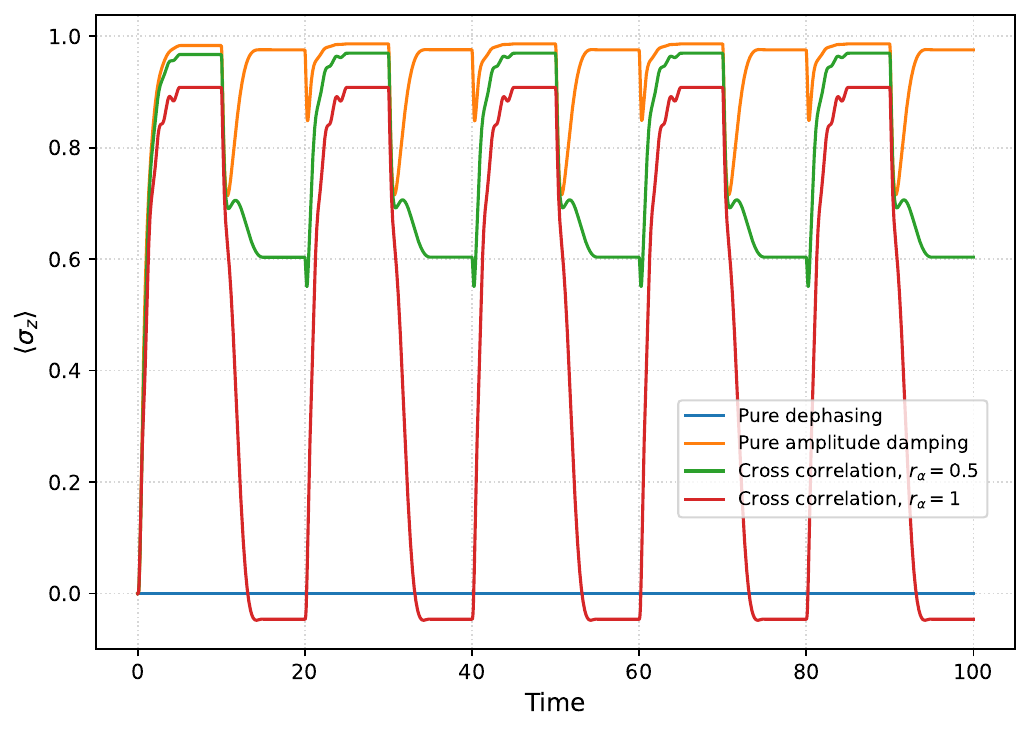}
\caption{
Time evolution of the population imbalance
$\langle\sigma_z\rangle$ of the working medium over successive Otto
cycles for the pure-dephasing and pure-amplitude-damping limits and
for longitudinal--transverse cross-spectral correlations with
$r_\alpha=0.5$ and $1$ ($\alpha=h,c$).
The other cross-spectral parameters are fixed at 
$\omega_{\rm corr}^{(\alpha)}=5$, and
$\phi_\alpha=\pi/2$, with cycle duration
$\tau_{\rm cyc}=20$.
All remaining parameters are fixed as specified in the text.
}
\label{fig:f7}
\end{figure}
Fig.~\ref{fig:f7} reveals a periodic population dynamics that is
strongly sensitive to the strength of the longitudinal--transverse
cross-spectral correlations in the reservoirs.  In the pure-dephasing limit,
$\langle\sigma_z\rangle$ remains constant, reflecting the absence of
population transfer and hence vanishing work output. Pure amplitude
damping induces only modest cycle-to-cycle population changes, whereas cross correlations produce increasingly large
oscillations as $r_\alpha$ is increased. For $r_\alpha=1$,
$\langle\sigma_z\rangle$ even becomes transiently negative, indicating
a temporary population inversion during part of the cycle.
For the convention $H_S(t)=-\omega(t)\sigma_z/2$, $\langle\sigma_z\rangle<0$
corresponds to a transient population inversion. 
Emergence of negative  $\langle\sigma_z\rangle$   at strong
cross correlations indicates that the reservoir can drive the working medium
far from the population distribution obtained in the absence of such correlations.
The increasing variation of $\langle\sigma_z\rangle$ with $r_\alpha$
enhances the energy contrast between the states prepared at the ends of
the hot and cold isochores, thereby providing a microscopic origin for
the corresponding enhancement of work output and power.

\section{Conclusions}
\label{sec:conclusion}

We have investigated the role of longitudinal--transverse cross-spectral
reservoir correlations in the finite-time thermodynamic performance of a
quantum Otto engine with a two-level working medium. The hot and cold
reservoirs were taken to be mutually independent, while the transverse and
longitudinal coupling channels within each reservoir were allowed to possess
finite cross correlations. These correlations were described through the
off-diagonal elements of a Hermitian positive-semidefinite spectral-density
matrix, thereby treating the auto- and cross-spectral contributions within a
common microscopic framework.\\

The reduced dynamics during the isochoric strokes was obtained within the
second-order time-convolutionless approach, retaining both dissipative and
reservoir-induced coherent contributions. In this description, the
cross-spectral terms modify the evolution of both populations and coherences
of the working medium. Their thermodynamic influence is therefore transferred
to the cycle through the state emerging from each isochore, which subsequently
serves as the initial state of the following unitary work stroke. Recasting the
TCL2 dynamics in Bloch-vector form allowed the four strokes to be combined into
a cycle map and provided a direct description of both the transient
cycle-to-cycle evolution and the asymptotic limit-cycle regime.\\

Our analysis shows that, even at fixed auto-spectral densities, the
off-diagonal reservoir spectrum provides an additional means of controlling
engine performance. Increasing the cross-spectral correlation strength can
enhance the extracted work and output power, while the magnitude of this
enhancement depends sensitively on the cross-spectral phase and characteristic
frequency scale. The correlations also modify the transient evolution of the
engine by changing the stroke-boundary states from cycle to cycle and,
consequently, the heat exchange, work output, transient work-to-heat ratio, and
power. In particular, they can alter the rate at which the engine approaches
its periodic steady state.\\

Once the limit cycle is established, the same sequence of stroke-boundary
states is reproduced in every cycle. For the population-preserving unitary
expansion and compression strokes considered here, this periodicity fixes the
asymptotic efficiency to the Otto value,
$\eta_{\rm LC}
=
1-\frac{\omega_c}{\omega_h}$.
The cross-spectral correlations therefore do not modify the limit-cycle
efficiency within the present thermodynamic convention. They nevertheless
continue to influence the limit-cycle populations and hence the heat exchanged,
the extracted work, and the asymptotic output power. The transient and
limit-cycle regimes thus display qualitatively different sensitivities to the
reservoir cross correlations: the former exhibits correlation-dependent
modifications of both the work-to-heat ratio and power, whereas in the latter
the principal effect survives through the work and power while the efficiency
is constrained to the Otto value.\\

These results show that longitudinal--transverse cross-spectral correlations provide an additional handle for controlling the finite-time power and transient thermodynamic response of the quantum Otto engine.
Thus, the off-diagonal reservoir spectra act as control parameters within a finite-time weak-coupling regime in which the TCL2 dynamics can retain non-Markovian memory effects through time-dependent dissipative and
Lamb-shift coefficients. Our results therefore identify cross-spectral reservoir correlations as a distinct reservoir-engineering resource for finite-time quantum thermal machine.

\bibliographystyle{JHEP}
\bibliography{ref_mod}

@article{Ithier2005,
  author = {Ithier, G. and Collin, E. and Joyez, P. and Meeson, P. J. and Vion, D. and Esteve, D. and Chiarello, F. and Shnirman, A. and Makhlin, Y. and Sch\"on, G.},
  title = {Decoherence in a superconducting quantum bit circuit},
  journal = {Physical Review B},
  volume = {72},
  pages = {134519},
  year = {2005},
  doi = {10.1103/PhysRevB.72.134519}
}

@article{Makhlin2001,
  author = {Makhlin, Y. and Sch\"on, G. and Shnirman, A.},
  title = {Quantum-state engineering with Josephson-junction devices},
  journal = {Reviews of Modern Physics},
  volume = {73},
  pages = {357--400},
  year = {2001},
  doi = {10.1103/RevModPhys.73.357}
}

@article{Paladino2014,
  author = {Paladino, E. and Galperin, Y. M. and Falci, G. and Altshuler, B. L.},
  title = {1/f noise: Implications for solid-state quantum information},
  journal = {Rev. Mod. Phys.},
  volume = {86},
  pages = {361--418},
  year = {2014},
  doi = {10.1103/RevModPhys.86.361}
}

@article{Witzel2006,
  author = {Witzel, W. M. and Das Sarma, S.},
  title = {Quantum theory for electron spin decoherence induced by nuclear spin dynamics},
  journal = {Physical Review B},
  volume = {74},
  pages = {035322},
  year = {2006},
  doi = {10.1103/PhysRevB.74.035322}
}

@incollection{NazirSchaller2018,
  author = {Nazir, Ahsan and Schaller, Gernot},
  title = {The Reaction Coordinate Mapping in Quantum Thermodynamics},
  booktitle = {Thermodynamics in the Quantum Regime},
  editor = {Binder, Felix and Correa, Luis A. and Gogolin, Christian and Anders, Janet and Adesso, Gerardo},
  publisher = {Springer},
  address = {Cham},
  pages = {551--577},
  year = {2018}
}

@article{Leggett1987,
  author = {Leggett, A. J. and Chakravarty, S. and Dorsey, A. T. and Fisher, M. P. A. and Garg, A. and Zwerger, W.},
  title = {Dynamics of the dissipative two-state system},
  journal = {Rev. Mod. Phys.},
  volume = {59},
  pages = {1--85},
  year = {1987},
  doi = {10.1103/RevModPhys.59.1}
}

@book{Breuer2002,
  author = {Breuer, Heinz-Peter and Petruccione, Francesco},
  title = {The Theory of Open Quantum Systems},
  publisher = {Oxford University Press},
  address = {Oxford},
  year = {2002}
}

@book{Weiss2012,
  author = {Weiss, Ulrich},
  title = {Quantum Dissipative Systems},
  edition = {4},
  publisher = {World Scientific},
  address = {Singapore},
  year = {2012}
}

@article{Caldeira1983,
  author    = {A. O. Caldeira and A. J. Leggett},
  title     = {Quantum Tunnelling in a Dissipative System},
  journal   = {Annals of Physics},
  volume    = {149},
  pages     = {374--456},
  year      = {1983}
}

@article{Clerk2010,
  author    = {A. A. Clerk and M. H. Devoret and S. M. Girvin and F. Marquardt and R. J. Schoelkopf},
  title     = {Introduction to Quantum Noise, Measurement, and Amplification},
  journal   = {Reviews of Modern Physics},
  volume    = {82},
  pages     = {1155--1208},
  year      = {2010}
}

@book{RivasHuelga2012,
  author    = {A. Rivas and S. F. Huelga},
  title     = {Open Quantum Systems: An Introduction},
  publisher = {Springer},
  year      = {2012}
}

@article{Palma1996,
  author  = {G. M. Palma and K.-A. Suominen and A. K. Ekert},
  title   = {Quantum computers and dissipation},
  journal = {Proceedings of the Royal Society of London. Series A:
             Mathematical, Physical and Engineering Sciences},
  volume  = {452},
  pages   = {567--584},
  year    = {1996}
}

@article{Jeske2013,
  author  = {Jan Jeske and Jared H. Cole},
  title   = {Derivation of Markovian master equations for spatially
             correlated decoherence},
  journal = {Physical Review A},
  volume  = {87},
  pages   = {052138},
  year    = {2013}
}

@article{FicekTanas2002,
  author  = {Z. Ficek and R. Tana{\'s}},
  title   = {Entangled states and collective nonclassical effects in
             two-atom systems},
  journal = {Physics Reports},
  volume  = {372},
  pages   = {369--443},
  year    = {2002}
}

@article{Krummheuer2002,
  author  = {B. Krummheuer and V. M. Axt and T. Kuhn},
  title   = {Theory of pure dephasing and the resulting absorption line
             shape in semiconductor quantum dots},
  journal = {Physical Review B},
  volume  = {65},
  pages   = {195313},
  year    = {2002}
}

@article{McCutcheonNazir2010,
  author  = {Dara P. S. McCutcheon and Ahsan Nazir},
  title   = {Quantum dot {Rabi} rotations beyond the weak
             exciton--phonon coupling regime},
  journal = {New Journal of Physics},
  volume  = {12},
  pages   = {113042},
  year    = {2010}
}

@article{Alicki1979,
  author  = {Alicki, Robert},
  title   = {The Quantum Open System as a Model of the Heat Engine},
  journal = {J. Phys. A: Math. Gen.},
  volume  = {12},
  number  = {5},
  pages   = {L103--L107},
  year    = {1979},
  doi     = {10.1088/0305-4470/12/5/007}
}

@article{Quan2007,
  author  = {Quan, H. T. and Liu, Yu-xi and Sun, C. P. and Nori, Franco},
  title   = {Quantum Thermodynamic Cycles and Quantum Heat Engines},
  journal = {Phys. Rev. E},
  volume  = {76},
  number  = {3},
  pages   = {031105},
  year    = {2007},
  doi     = {10.1103/PhysRevE.76.031105}
}

@article{Kosloff2013,
  author  = {Kosloff, Ronnie},
  title   = {Quantum Thermodynamics: A Dynamical Viewpoint},
  journal = {Entropy},
  volume  = {15},
  number  = {6},
  pages   = {2100--2128},
  year    = {2013},
  doi     = {10.3390/e15062100}
}

@article{Kieu2004,
  author    = {Kieu, T. D.},
  title     = {The Second Law, Maxwell's Demon, and Work Derivable from
               Quantum Heat Engines},
  journal   = {Physical Review Letters},
  volume    = {93},
  number    = {14},
  pages     = {140403},
  year      = {2004},
  publisher = {American Physical Society},
  doi       = {10.1103/PhysRevLett.93.140403}
}

@article{Peterson2019,
  author    = {Peterson, John P. S. and Batalh{\~a}o, Tiago B. and
               Herrera, Marcela and Souza, Alexandre M. and
               Sarthour, Roberto S. and Oliveira, Ivan S. and
               Serra, Roberto M.},
  title     = {Experimental Characterization of a Spin Quantum Heat Engine},
  journal   = {Physical Review Letters},
  volume    = {123},
  number    = {24},
  pages     = {240601},
  year      = {2019},
  publisher = {American Physical Society},
  doi       = {10.1103/PhysRevLett.123.240601}
}

@article{Brandner2017,
  author  = {Brandner, Kay and Bauer, Michael and Seifert, Udo},
  title   = {Universal Coherence-Induced Power Losses of Quantum
             Heat Engines in Linear Response},
  journal = {Phys. Rev. Lett.},
  volume  = {119},
  number  = {17},
  pages   = {170602},
  year    = {2017},
  doi     = {10.1103/PhysRevLett.119.170602}
}

@article{Camati2019,
  author  = {Camati, Patrice A. and Santos, Jonas F. G. and
             Serra, Roberto M.},
  title   = {Coherence Effects in the Performance of the Quantum
             Otto Heat Engine},
  journal = {Phys. Rev. A},
  volume  = {99},
  number  = {6},
  pages   = {062103},
  year    = {2019},
  doi     = {10.1103/PhysRevA.99.062103}
}

@article{Rossnagel2014,
  author  = {Ro{\ss}nagel, Johannes and Abah, Obinna and
             Schmidt-Kaler, Ferdinand and Singer, Kilian and Lutz, Eric},
  title   = {Nanoscale Heat Engine Beyond the Carnot Limit},
  journal = {Phys. Rev. Lett.},
  volume  = {112},
  number  = {3},
  pages   = {030602},
  year    = {2014},
  doi     = {10.1103/PhysRevLett.112.030602}
}

@article{Mendonca2020,
  author  = {Mendon{\c{c}}a, Taysa M. and Souza, Alexandre M. and
             de Assis, Rog{\'e}rio J. and de Almeida, Norton G. and
             Sarthour, Roberto S. and Oliveira, Ivan S. and
             Villas-Boas, Celso J.},
  title   = {Reservoir Engineering for Maximally Efficient
             Quantum Engines},
  journal = {Phys. Rev. Research},
  volume  = {2},
  number  = {4},
  pages   = {043419},
  year    = {2020},
  doi     = {10.1103/PhysRevResearch.2.043419}
}

@article{Newman2017,
  author  = {Newman, David and Mintert, Florian and Nazir, Ahsan},
  title   = {Performance of a Quantum Heat Engine at Strong
             Reservoir Coupling},
  journal = {Phys. Rev. E},
  volume  = {95},
  number  = {3},
  pages   = {032139},
  year    = {2017},
  doi     = {10.1103/PhysRevE.95.032139}
}

@article{Shirai2021,
  author  = {Shirai, Yuji and Hashimoto, Kazunari and Tezuka, Ryuta and
             Uchiyama, Chikako and Hatano, Naomichi},
  title   = {Non-Markovian Effect on Quantum Otto Engine:
             Role of System--Reservoir Interaction},
  journal = {Phys. Rev. Research},
  volume  = {3},
  number  = {2},
  pages   = {023078},
  year    = {2021},
  doi     = {10.1103/PhysRevResearch.3.023078}
}

@article{Ptaszynski2022,
  author  = {Ptaszy{\'n}ski, Krzysztof},
  title   = {Non-Markovian Thermal Operations Boosting the
             Performance of Quantum Heat Engines},
  journal = {Phys. Rev. E},
  volume  = {106},
  number  = {1},
  pages   = {014114},
  year    = {2022},
  doi     = {10.1103/PhysRevE.106.014114}
}

@article{PazSilva2019,
  author  = {Paz-Silva, Gerardo A. and Norris, Leigh M. and
             Beaudoin, F{\'e}lix and Viola, Lorenza},
  title   = {Extending Comb-Based Spectral Estimation to
             Multiaxis Quantum Noise},
  journal = {Phys. Rev. A},
  volume  = {100},
  number  = {4},
  pages   = {042334},
  year    = {2019},
  doi     = {10.1103/PhysRevA.100.042334}
}

@article{Dutta2026,
  author        = {Dutta, Siddhartha and Mondal, Sujay and
                   Bandyopadhyay, Abhijit},
  title         = {Effect of Cross-Spectral Correlations on Qubit Dynamics:
                   Coherence Revival and Relaxation Modulation},
  journal       = {arXiv preprint},
  year          = {2026},
  eprint        = {2608.04672},
  archivePrefix = {arXiv},
  primaryClass  = {quant-ph},
  doi           = {10.48550/arXiv.2608.04672}
}

@article{Talkner2009,
  author  = {Talkner, Peter and Campisi, Michele and H{\"a}nggi, Peter},
  title   = {Fluctuation Theorems in Driven Open Quantum Systems},
  journal = {Journal of Statistical Mechanics: Theory and Experiment},
  volume  = {2009},
  number  = {02},
  pages   = {P02025},
  year    = {2009},
  doi     = {10.1088/1742-5468/2009/02/P02025}
}

@article{PerarnauLlobet2018,
  author  = {Perarnau-Llobet, Mart{\'i} and Wilming, Henrik and
             Riera, Arnau and Gallego, Rodrigo and Eisert, Jens},
  title   = {Strong Coupling Corrections in Quantum Thermodynamics},
  journal = {Physical Review Letters},
  volume  = {120},
  number  = {12},
  pages   = {120602},
  year    = {2018},
  doi     = {10.1103/PhysRevLett.120.120602}
}

@article{FeynmanVernon1963,
  author  = {Feynman, Richard P. and Vernon, Frank L.},
  title   = {The Theory of a General Quantum System Interacting with a Linear Dissipative System},
  journal = {Annals of Physics},
  volume  = {24},
  pages   = {118--173},
  year    = {1963},
  doi     = {10.1016/0003-4916(63)90068-X}
}

@article{CaldeiraLeggett1983,
  author  = {Caldeira, A. O. and Leggett, A. J.},
  title   = {Path Integral Approach to Quantum Brownian Motion},
  journal = {Physica A: Statistical Mechanics and its Applications},
  volume  = {121},
  number  = {3},
  pages   = {587--616},
  year    = {1983},
  doi     = {10.1016/0378-4371(83)90013-4}
}

@article{Shibata1977,
  author  = {Shibata, Fumiaki and Takahashi, Yoshinori and Hashitsume, Natsuki},
  title   = {A Generalized Stochastic Liouville Equation. Non-Markovian versus Memoryless Master Equations},
  journal = {Journal of Statistical Physics},
  volume  = {17},
  number  = {4},
  pages   = {171--187},
  year    = {1977},
  doi     = {10.1007/BF01040100}
}

@article{BreuerKappler2001,
  author  = {Breuer, Heinz-Peter and Kappler, Bernd and Petruccione, Francesco},
  title   = {The Time-Convolutionless Projection Operator Technique in the Quantum Theory of Dissipation and Decoherence},
  journal = {Annals of Physics},
  volume  = {291},
  number  = {1},
  pages   = {36--70},
  year    = {2001},
  doi     = {10.1006/aphy.2001.6152}
}

@article{ChaturvediShibata1979,
  author  = {Chaturvedi, S. and Shibata, F.},
  title   = {Time-Convolutionless Projection Operator Formalism for Elimination of Fast Variables. Applications to Brownian Motion},
  journal = {Zeitschrift f{\"u}r Physik B: Condensed Matter},
  volume  = {35},
  number  = {3},
  pages   = {297--308},
  year    = {1979},
  doi     = {10.1007/BF01319852}
}

@article{deAssis2019,
  author  = {de Assis, Rog{\'e}rio J. and de Mendon{\c c}a, Taysa M.
             and Villas-Boas, Celso J. and de Souza, Alexandre M.
             and Sarthour, Roberto S. and Oliveira, Ivan S.
             and de Almeida, Norton G.},
  title   = {Efficiency of a Quantum Otto Heat Engine Operating under
             a Reservoir at Effective Negative Temperatures},
  journal = {Phys. Rev. Lett.},
  volume  = {122},
  number  = {24},
  pages   = {240602},
  year    = {2019},
  doi     = {10.1103/PhysRevLett.122.240602}
}

@article{Khan2024,
  author  = {Khan, Muhammad Qasim and Dong, Wenzheng and
             Norris, Leigh M. and Viola, Lorenza},
  title   = {Multiaxis Quantum Noise Spectroscopy Robust to Errors
             in State Preparation and Measurement},
  journal = {Phys. Rev. Applied},
  volume  = {22},
  number  = {2},
  pages   = {024074},
  year    = {2024},
  doi     = {10.1103/PhysRevApplied.22.024074}
}

@article{Gustavsson2011,
  author  = {Gustavsson, Simon and Bylander, Jonas and Yan, Fei and
             Oliver, William D. and Yoshihara, Fumiki and Nakamura, Yasunobu},
  title   = {Noise Correlations in a Flux Qubit with Tunable Tunnel Coupling},
  journal = {Phys. Rev. B},
  volume  = {84},
  number  = {1},
  pages   = {014525},
  year    = {2011},
  doi     = {10.1103/PhysRevB.84.014525}
}

@article{vonLupke2020,
  author  = {von L{\"u}pke, Uwe and Beaudoin, F{\'e}lix and
             Norris, Leigh M. and Sung, Youngkyu and Winik, Roni and
             Qiu, Jack Y. and Kjaergaard, Morten and Kim, David and
             Yoder, Jonilyn and Gustavsson, Simon and Viola, Lorenza and
             Oliver, William D.},
  title   = {Two-Qubit Spectroscopy of Spatiotemporally Correlated
             Quantum Noise in Superconducting Qubits},
  journal = {PRX Quantum},
  volume  = {1},
  number  = {1},
  pages   = {010305},
  year    = {2020},
  doi     = {10.1103/PRXQuantum.1.010305}
}

@article{RojasArias2026,
  author  = {Rojas-Arias, J. S. and Stano, P. and Wu, Y.-H. and
             Camenzind, L. C. and Tarucha, S. and Loss, D.},
  title   = {Noise Cross-Correlations from Single-Shot Measurements},
  journal = {PRX Quantum},
  volume  = {7},
  number  = {2},
  pages   = {020351},
  year    = {2026},
  doi     = {10.1103/1m1q-rvth}
}

@article{Sun2025,
  author  = {Sun, Ke and Kang, Mingyu and Nuomin, Hanggai and Schwartz, George
             and Beratan, David N. and Brown, Kenneth R. and Kim, Jungsang},
  title   = {Quantum Simulation of Spin-Boson Models with Structured Bath},
  journal = {Nature Communications},
  volume  = {16},
  pages   = {4042},
  year    = {2025},
  doi     = {10.1038/s41467-025-59296-y}
}

@book{BreuerPetruccione2002,
  author    = {Breuer, Heinz-Peter and Petruccione, Francesco},
  title     = {The Theory of Open Quantum Systems},
  publisher = {Oxford University Press},
  address   = {Oxford},
  year      = {2002},
  isbn      = {9780198520634}
}

@article{FarinaGiovannetti2019,
  author  = {Farina, Donato and Giovannetti, Vittorio},
  title   = {Open-quantum-system dynamics: Recovering positivity of the Redfield equation via the partial secular approximation},
  journal = {Physical Review A},
  volume  = {100},
  number  = {1},
  pages   = {012107},
  year    = {2019},
  doi     = {10.1103/PhysRevA.100.012107}
}

@article{RivasHuelgaPlenio2014,
  author  = {Rivas, {\'A}ngel and Huelga, Susana F. and Plenio, Martin B.},
  title   = {Quantum non-Markovianity: characterization, quantification and detection},
  journal = {Reports on Progress in Physics},
  volume  = {77},
  number  = {9},
  pages   = {094001},
  year    = {2014},
  doi     = {10.1088/0034-4885/77/9/094001}
}

@article{deVegaAlonso2017,
  author  = {de Vega, In{\'e}s and Alonso, Daniel},
  title   = {Dynamics of non-Markovian open quantum systems},
  journal = {Reviews of Modern Physics},
  volume  = {89},
  number  = {1},
  pages   = {015001},
  year    = {2017},
  doi     = {10.1103/RevModPhys.89.015001}
}

\end{document}